\documentclass[useAMS,usenatbib,twocolumn]{mnras}
\usepackage{graphicx}
\usepackage{amsmath}
\usepackage{amssymb}
\usepackage{xspace}
\usepackage{commath}
\usepackage{times}
\usepackage{bm} 
\usepackage{balance}
\usepackage{hyperref}
\usepackage{mathtools}
\usepackage{stmaryrd}

\hypersetup{dvips, colorlinks=true, linkcolor=black, citecolor=black, filecolor=blue, urlcolor=black}

\usepackage{acronym}
\newacro{DF}{distribution function}
\newacroplural{DFs}{distribution functions}
\newcommand{\DF}{\ac{DF}}

\newacro{GL}{Gauss--Legendre}
\newcommand{\GL}{\ac{GL}}
\newacro{SG}{Savitzky--Golay}
\newcommand{\SG}{\ac{SG}}

\newcommand{\rc}{\mathrm{c}}
\newcommand{\rd}{\mathrm{d}}
\newcommand{\re}{\mathrm{e}}
\newcommand{\ri}{\mathrm{i}}
\newcommand{\p}{\partial}
\newcommand{\half}{\tfrac{1}{2}}
\newcommand{\bM}{\mathbf{M}}
\newcommand{\bn}{\mathbf{n}}
\newcommand{\bT}{\bm{\theta}}
\newcommand{\bJ}{\mathbf{J}}
\newcommand{\bO}{\mathbf{\Omega}}
\newcommand{\umax}{u_{\mathrm{max}}}
\newcommand{\mL}{\mathcal{L}}
\newcommand{\intL}{\int_{\substack{ \\[0.83ex] \hphantom{-1} \\ \hspace{-1.9em} \mL}}^{\hphantom{1}} \!\!}
\newcommand{\intLb}{\int_{\substack{ \\[0.83ex] -1 \\ \hspace{-1.9em} \mL}}^{1}}
\newcommand{\intLbb}{\int_{\substack{ \\[0.83ex] 0 \\ \hspace{-1.2em} \mL}}^{1}}
\newcommand{\intLinf}{\int_{\substack{ \\[0.83ex] - \infty \\ \hspace{-2.3em} \mL}}^{+ \infty}}
\newcommand{\eps}{\epsilon}
\newcommand{\veps}{\varepsilon}
\newcommand{\Ku}{K_{u}}
\newcommand{\Kv}{K_{v}}
\newcommand{\omegaR}{\omega_{\mathrm{R}}}
\newcommand{\omegaM}{\omega_{\mathrm{M}}}
\newcommand{\Ra}{R_{\mathrm{a}}}
\newcommand{\RePart}{\mathrm{Re}}
\newcommand{\ImPart}{\mathrm{Im}}
\newcommand{\Pc}{P_{\mathrm{c}}}
\newcommand{\dB}{\mathrm{dB}}
\newcommand{\tmax}{t_{\mathrm{max}}}
\newcommand{\HU}{\mathrm{HU}}
\newcommand{\Nreal}{N_{\mathrm{real}}}
\newcommand{\br}{\mathbf{r}}

\newcommand{\bv}{\mathbf{v}}
\newcommand{\bISO}{b_{\mathrm{c}}}
\newcommand{\nmax}{n_{\mathrm{max}}}
\newcommand{\Rb}{R_{\mathrm{b}}}
\newcommand{\bI}{\mathbf{I}}
\newcommand{\mP}{\mathcal{P}}
\newcommand{\bN}{\mathbf{N}}
\newcommand{\vtheta}{\vartheta}
\newcommand{\Ftot}{F_{\mathrm{tot}}}
\newcommand{\Leg}{\mathrm{Leg}}
\newcommand{\cst}{\mathrm{cst.}}
\newcommand{\Kc}{K_{\mathrm{c}}}
\newcommand{\Min}{\mathrm{Min}}

\newcommand{\xc}{x_{\mathrm{c}}}
\newcommand{\yc}{y_{\mathrm{c}}}
\newcommand{\zc}{z_{\mathrm{c}}}
\newcommand{\TM}{T_{\mathrm{M}}}
\newcommand{\Tc}{T_{\mathrm{c}}}
\newcommand{\Mf}{M_{\mathrm{f}}}
\newcommand{\Nf}{N_{\mathrm{f}}}
\newcommand{\oxc}{\overline{x}_{\mathrm{c}}}
\newcommand{\Rc}{R_{\mathrm{c}}}
\newcommand{\hdR}{\widehat{\delta R}}
\newcommand{\Nboot}{N_{\mathrm{boot}}}
\newcommand{\alphac}{\alpha_{\mathrm{c}}}
\newcommand{\betac}{\beta_{\mathrm{c}}}
\newcommand{\vnm}{v_{\mathbf{n}}^{-}}
\newcommand{\vnp}{v_{\mathbf{n}}^{+}}
\newcommand{\vphi}{\varphi}
\newcommand{\kp}{k^{\prime}}

\newcommand{\Rv}{R_{\mathrm{v}}}
\newcommand{\bzero}{\mathbf{0}}
\newcommand{\mC}{\mathcal{C}}
\newcommand{\rperi}{r_{\mathrm{p}}}
\newcommand{\rapo}{r_{\mathrm{a}}}

\begin{document}

\title[Damped modes of stellar clusters]{Linear response theory and damped modes of stellar clusters}

\author[J.-B. Fouvry \& S. Prunet]{Jean-Baptiste Fouvry$^{1}$ and Simon Prunet$^{2}$\\
\noindent
$^{1}$ CNRS and Sorbonne Universit\'e, UMR 7095, Institut d'Astrophysique de Paris, 98 bis Boulevard Arago, F-75014 Paris, France\\
$^{2}$ Laboratoire Lagrange, Universit\'e C\^ote d'Azur, Observatoire de la C\^ote d'Azur, CNRS, Parc Valrose, 06104 Nice Cedex 2, France
}

\maketitle

\begin{abstract}
Because all stars contribute to its gravitational potential,
stellar clusters amplify perturbations collectively.
In the limit of small fluctuations, this is described
through linear response theory,
via the so-called response matrix.
While the evaluation of this matrix is somewhat straightforward for unstable modes (i.e.\ with a positive growth rate),
it requires a careful analytic continuation
for damped modes
(i.e.\ with a negative growth rate).
We present a generic method to perform
such a calculation in spherically symmetric stellar clusters.
When applied to an isotropic isochrone cluster,
we recover the presence of a low-frequency weakly damped
${ \ell \!=\! 1 }$ mode.
We finally use a set of direct $N$-body simulations
to test explicitly this prediction
through the statistics of the correlated random walk
undergone by a cluster's density centre.
\end{abstract}


\section{Introduction}
\label{sec:intro}

All the stars in a stellar cluster contribute
collectively to the system's potential.
This self-consistency naturally allows the cluster
to respond and amplify disturbances.
In the limit of small perturbations,
this is described by linear response
theory~\citep[see, e.g.\@,][]{BinneyTremaine2008}.
Such a generic machinery is paramount to
characterise a cluster's possible unstable modes,
i.e.\ with an amplitude growing exponentially in time,
for example via the radial-orbit instability
in radially anisotropic stellar
clusters~\citep[see, e.g.\@,][for reviews]{Merritt1999,Marechal+2011}.

Yet, even if a cluster is dynamically stable,
i.e.\ there are no unstable modes,
it does not imply that it is dynamically static.
Indeed, the cluster can still sustain
damped modes, i.e.\ with an amplitude
decaying exponentially in time.
These are commonly called Landau damped modes,
and require a careful analytic continuation
of the response matrix following Landau's prescription.
While these modes naturally damp when in isolation,
they can also be excited
by external perturbations,
e.g.\@, fly-bys~\citep{Weinberg1989}
or continuously seeded by the cluster's intrinsic
thermal Poisson shot noise~\citep{Weinberg1998}.

The importance of these damped modes
in globular clusters is the clearest through its ${ \ell \!=\! 1 }$
`sloshing' mode that induces slow long-lasting oscillations
of a cluster's centre~\citep[see, e.g.\@, ][and references therein]{Heggie+2020},
and is also imprinted through the strong amplification
of large-scale dipole fluctuations~\citep{Weinberg1998,Hamilton+2018,Lau+2019}.
In a seminal work, \cite{Weinberg1994} developed
an elegant numerical procedure
to systematically evaluate a cluster's response matrix
for such damped frequencies.
This is the problem that we revisit here
putting forward a different approach
to perform efficiently and explicitly this analytic continuation.

The present paper is organised as follows.
In~\S\ref{sec:LinearResponseTheory},
we briefly review the linear response theory
of stellar systems.
In~\S\ref{sec:Analytic continuation},
we rewrite the cluster's response matrix
to perform its analytic continuation
necessary to capture the damped part
of the modes' spectrum.
We apply this method in~\S\ref{sec:Application}
to the isotropic isochrone cluster
and recover the presence of a weakly damped
${ \ell \!=\! 1 }$ mode therein,
which is subsequently compared
with direct $N$-body simulations.
Finally, we conclude in~\S\ref{sec:conclusion}.
In all these sections, technical details are either deferred
to Appendices or to appropriate references.

\section{Linear response theory}
\label{sec:LinearResponseTheory}

The linear response theory
of an integrable long-range interacting system
of dimension $d$
is generically characterised by its response matrix,
${ \bM (\omega) }$~\citep[see \S{5.3} in][]{BinneyTremaine2008},
reading
\begin{equation}
M_{\alpha \beta} (\omega) = - (2 \pi)^{d} \sum_{\bn \in \mathbb{Z}^{d}} \intL \!\! \rd \bJ \, \frac{\bn \!\cdot\! \p \Ftot / \p \bJ}{\bn \!\cdot\! \bO (\bJ) - \omega} \, \psi^{(\alpha) *}_{\bn} (\bJ) \, \psi^{(\beta)}_{\bn} (\bJ) .
\label{def_bM}
\end{equation}
In that expression, we introduced angle-action coordinates
(of dimension $d$) as ${ (\bT , \bJ) }$,
and the system's quasi-stationary \DF\@, ${ \Ftot (\bJ) }$,
normalised so that ${ \! \int \! \rd \bT \rd \bJ \Ftot \!=\! M }$,
with $M$ the system's total active mass.
Equation~\eqref{def_bM} contains the orbital frequencies, ${ \bO (\bJ) }$,
as well as the Fourier resonance numbers ${ \bn \!\in\! \mathbb{Z}^{d} }$.
Finally, following the basis method (see~\S\ref{sec:LinearTheory3D}),
Eq.~\eqref{def_bM} also involves a set
of biorthogonal potential basis elements, ${ \psi^{(\alpha)} }$.

A system sustains a mode at the (complex) frequency $\omega$,
if one has ${ \det [\bI \!-\! \bM (\omega) ] \!=\! 0 }$,
with $\bI$ the identity matrix.
Here, the sign of ${ \ImPart[\omega] }$
controls the nature of the mode,
as ${ \ImPart[\omega] \!>\! 0 }$ corresponds to unstable modes,
${ \ImPart[\omega] \!=\! 0 }$ to neutral modes,
and ${ \ImPart[\omega] \!<\! 0 }$ to (Landau) damped ones.
We refer to~\cite{Case1959,Lee2018,Polyachenko+2021,Lau+2021}
and references therein
for detailed discussions on the subtle distinction
between genuine (van Kampen) modes
and the present Landau damped modes.

Importantly, for ${ \ImPart [\omega] \!<\! 0 }$
the action integral from Eq.~\eqref{def_bM}
must be computed using Landau's prescription~\citep[see, e.g.\@, \S{5.2.4} in][]{BinneyTremaine2008},
hence the notation ${ \!\int_{\mathcal{L}} \rd \bJ }$.
In the case of ${1D}$ homogeneous system,
this prescription takes the simple form
\begin{equation}
\intLinf \!\!\!\!\!\!\!\! \rd u \, \frac{G (u)}{u \!-\! \omega} =
\begin{cases}
\displaystyle \!\! \int_{- \infty}^{+ \infty} \!\!\!\!\!\!\!\! \rd u \, \frac{G (u)}{u \!-\! \omega} & \!\!\! \text{if} \;\; \ImPart[\omega] > 0 ,
\\[2.0ex]
\displaystyle \mP \!\! \int_{- \infty}^{+ \infty} \!\!\!\!\!\!\!\! \rd u \, \frac{G (u)}{u \!-\! \omega} + \ri \pi \, G (\omega) & \!\!\! \text{if} \;\; \ImPart[\omega] = 0 ,
\\[2.0ex]
\displaystyle \!\! \int_{- \infty}^{+ \infty} \!\!\!\!\!\!\!\! \rd u \, \frac{G (u)}{u - \omega} + 2 \ri \pi \, G (\omega) & \!\!\! \text{if} \;\; \ImPart[\omega] < 0 
\end{cases}
\label{Landau_Prescription}
\end{equation}
where the function ${ u \!\mapsto\! G (u) }$
is assumed to be analytic,
e.g.\@, ${ G (u) \!\propto\! u \, \re^{-u^{2}} }$
in the case of a Maxwellian distribution.
In Eq.~\eqref{Landau_Prescription},
we also introduced $\mP$ as Cauchy's principal value.
Our goal is to illustrate how one may adapt
Landau's prescription to the case of self-gravitating systems.

We now focus our interest
on spherically symmetric stellar systems.
In that case, as detailed in~\S{4} of~\cite{Hamilton+2018},
the response matrix from Eq.~\eqref{def_bM}
decouples the various spherical harmonics ${ \ell \!\geq\! 0 }$
from one another.
The response matrix is then limited
to a ${2D}$ action integral.
For a given harmonic $\ell$,
Eq.~\eqref{def_bM} generically takes the form
\begin{equation}
M^{\ell}_{pq} (\omega) = \sum_{\bn \in \mathbb{Z}^{2}} \intL \!\! \rd \bJ \, \frac{G_{pq}^{\ell \bn} (\bJ)}{\bn \!\cdot\! \bO (\bJ) - \omega} ,
\label{shape_bM}
\end{equation}
with ${ (p,q) }$ the radial indices associated
with the basis decomposition,
and with ${ G_{pq}^{\ell \bn} (\bJ) }$
fully spelled out in Eq.~\eqref{def_Gpq}.
In Eq.~\eqref{shape_bM},
we introduced the ${ 2D }$ action coordinates
${ \bJ \!=\! (J_{r} , L) }$,
as the radial action and angular momentum.
Similarly, the resonance vectors are ${2D}$,
i.e.\ ${ \bn \!=\! (n_{1} , n_{2}) }$,
with the orbital frequencies
${ \bO (\bJ) \!=\! (\Omega_{1} (\bJ) , \Omega_{2} (\bJ)) }$,
respectively associated with the radial and azimuthal motions.
\linebreak

There are two main difficulties associated
with the application of Landau's prescription
to Eq.~\eqref{shape_bM}:
\begin{itemize}
\item[(A)] It involves the resonant denominator,
${ 1 / (\bn \!\cdot\! \bO (\bJ) \!-\! \omega) }$,
considerably more intricate than the ${ 1 / (u \!-\! \omega) }$
appearing in the homogeneous case.
Orbital frequencies are non-trivial functions
of the action variables,
i.e.\ non-trivial functions of the coordinate
w.r.t.\ which the integrals are performed.
In that sense, the stellar resonance condition
is not `aligned' with one of the integration coordinates.
\\[-1.0ex]
\item[(B)] The numerator, ${ G_{pq}^{\ell \bn} (\bJ) }$,
is an expensive numerical function
which, as such, cannot be evaluated
for arbitrary complex arguments.
This is in stark constrast with the homogeneous case,
where it is generically assumed that the numerator, ${ G (u) }$,
is an explicitly known analytic function.
\end{itemize}
Our goal is now to circumvent these two issues
to explicitly apply Landau's prescription to the stellar case.

\section{Analytic continuation}
\label{sec:Analytic continuation}

In order to simplify the notations,
we now drop the dependence w.r.t.\
the harmonics $\ell$,
as well as w.r.t.\ the considered basis elements ${ (p,q) }$.
As a consequence, following Eq.~\eqref{shape_bM},
our goal is to evaluate an expression of the form
\begin{equation}
M_{\bn} (\omega) = \intL \!\! \rd \bJ \, \frac{G_{\bn} (\bJ)}{\bn \!\cdot\! \bO (\bJ) - \omega} ,
\label{shape_bM_short}
\end{equation}
for a given resonance vector ${ \bn \!=\! (n_{1} , n_{2}) }$.

We now assume that the cluster's density
follows an outward decreasing core profile,
e.g.\@ as in the isochrone cluster.
As such, we introduce the natural frequency scale
\begin{equation}
\Omega_{0} = \Omega_{1} \big( J_{r} \!\to\! 0 , L \!\to\! 0 \big) ,
\label{def_Omega0}
\end{equation}
corresponding to the frequency of harmonic oscillation
in the cluster's very core.
In order to `align' the resonant denominator from Eq.~\eqref{shape_bM_short},
we introduce the new dimensionless coordinates
\begin{equation}
\alpha = \frac{\Omega_{1}}{\Omega_{0}} ;
\quad
\beta = \frac{\Omega_{2}}{\Omega_{1}} ,
\label{def_alpha_beta}
\end{equation}
so that $\alpha$ corresponds to the (dimensionless) radial frequency,
and ${ \beta }$ to the ratio of the azimuthal and radial frequencies.
Such a choice stems from the fact that $\alpha$ and $\beta$
are naturally obtained through the angle-action coordinates mapping~\citep{TremaineWeinberg1984} --
see also~\S\ref{sec:LinearTheory3D}.
Importantly, as long as the cluster's potential is not degenerate,
e.g.\@, different from the Keplerian and harmonic one,
the mapping ${ \bJ \!\mapsto\! (\alpha , \beta) }$ is bijective.
As such, an orbit can unambiguously
be characterised by its two orbital frequencies.
In addition, we note that $\alpha$ and $\beta$ satisfy the range constraints
\begin{equation}
0 \leq \alpha \leq 1 ;
\quad
 \half \leq \beta \leq \betac (\alpha) .
\label{ranges_alpha_beta}
\end{equation}
Here, ${ \alpha \!\to\! 0 }$ corresponds the outer regions of the cluster,
while ${ \alpha \!\to\! 1 }$ corresponds to the inner regions.
One has ${ \beta \!=\! \half }$ along radial orbits (i.e.\ ${ J_{r} \!\to\! + \infty }$),
while ${ \beta \!=\! \betac (\alpha) }$ stands for the value
of the frequency ratio along circular orbits (i.e.\ ${ J_{r} \!\to\! 0 }$).
We refer to Eq.~\eqref{def_betac} for an explicit expression
of ${ \betac (\alpha) }$ in the case of the isochrone potential.

All in all, we may rewrite Eq.~\eqref{shape_bM_short} as
\begin{equation}
M_{\bn} (\omega) = \intLbb \!\! \rd \alpha \!\! \int_{\half}^{\betac (\alpha)} \!\! \rd \beta \, \frac{G_{\bn} (\alpha , \beta)}{n_{1} \, \alpha + n_{2} \, \alpha \, \beta - \omega} ,
\label{bM_alpha_beta}
\end{equation}
where we introduced
${ G_{\bn} (\alpha , \beta) \!=\! G_{\bn} (\bJ) \, |\p \bJ / \p (\alpha , \beta) |/\Omega_{0} }$,
and make the convenient replacement
${ \omega/\Omega_{0} \!\to\! \omega }$
for the rest of this section.
We refer to~\S\ref{sec:IsochronePotential}
for an explicit expression of the Jacobian
of that transformation for the isochrone potential.

The next step of the calculation
is to perform one additional change of variables
so as to fully align the resonant denominator
from Eq.~\eqref{bM_alpha_beta}
with one of the integration variables.
To proceed forward, for a given resonance vector $\bn$,
we introduce the (real) resonance frequency
\begin{equation}
\omega_{\bn} (\alpha , \beta) = n_{1} \, \alpha + n_{2} \, \alpha \, \beta .
\label{def_omegan}
\end{equation}
In practice, we then perform
a change of variables of the form
${ (\alpha , \beta) \!\mapsto\! (u , v) }$,
so that the new coordinates
satisfy the three constraints:
(i) ${ u \!\propto\! \omega_{\bn} \!+\! \cst }$;
(ii) ${ -1 \!\leq\! u \!\leq\! 1 }$;
(iii) ${ \vnm (u) \!\leq\! v \!\leq\! \vnp (u) }$.
We spell out explicitly this change of variables
in~\S\ref{sec:OrbitalFrequencies},
and tailor it for the isochrone case
in~\S\ref{sec:IsochronePotential}.

All in all, Eq.~\eqref{bM_alpha_beta} can then
be rewritten under the generic form
\begin{equation}
M_{\bn} (\omega) = \!\! \intLb \!\! \rd u \! \int_{\vnm(u)}^{\vnp (u)} \!\!\!\! \rd v \, \, \frac{G_{\bn} (u,v)}{u - \varpi_{\bn} (\omega)} ,
\label{expression_M}
\end{equation}
with the detailed expression of ${ G_{\bn} (u , v) }$
given in Eq.~\eqref{exp_G_App}.
In Eq.~\eqref{expression_M}, we also introduced
the rescaled (complex) frequency
\begin{equation}
\varpi_{\bn} (\omega) = \frac{2 \omega - \omega_{\bn}^{\max} - \omega_{\bn}^{\min}}{\omega_{\bn}^{\max} - \omega_{\bn}^{\min}} ,
\label{def_varpi}
\end{equation}
with ${ \omega_{\bn}^{\min} \!=\! \Min_{(\alpha,\beta)}[\omega_{\bn} (\alpha , \beta)] }$,
and similarly for ${ \omega_{\max} }$.
Given that $\omega_{\bn}^{\min}$ and $\omega_{\bn}^{\max}$
are both real, we emphasise from Eq.~\eqref{def_varpi}
that $\omega$ and ${ \varpi_{\bn} (\omega) }$
share the same sign for their imaginary part,
so that Landau's prescription (Eq.~\eqref{Landau_Prescription})
also naturally applies with ${ \varpi_{\bn} (\omega) }$.

At this stage, we have made great progress,
as the problem (A) from~\S\ref{sec:LinearResponseTheory}
has been fully circumvented in Eq.~\eqref{expression_M}.
Indeed, the resonant denominator in that equation
now takes the simple form ${ 1/ (u \!-\! \varpi_{\bn}) }$,
with $u$ one of the integration variable.

It now only remains to deal
with the problem (B) from~\S\ref{sec:LinearResponseTheory},
namely to perform the analytic continuation
of Eq.~\eqref{expression_M}.
In order to further simplify the notations,
we now drop the dependences w.r.t.\ the resonance vector $\bn$,
and the integral ${ \!\int\! \rd v }$,
and make the replacement ${ \varpi_{\bn} (\omega) \!\to\! \omega }$.
As a consequence,
Eq.~\eqref{expression_M} generically asks
for the computation of an expression of the form
\begin{equation}
M (\omega) = \!\! \intLb \!\! \rd u \, \frac{G (u)}{u - \omega} ,
\label{def_I}
\end{equation}

To proceed forward, we follow the same approach as~\cite{Robinson1990}
by projecting the function ${ u \!\mapsto\! G (u) }$
onto an explicit and analytic basis function.
Considering that the integration range from Eq.~\eqref{def_I}
is finite, a natural choice is Legendre polynomials,
${ P_{k} (u) }$.
More precisely, given a maximum order $\Ku$,
we write the expansion
\begin{equation}
G (u) = \sum_{k = 0}^{\Ku - 1} a_{k} \, P_{k} (u) ,
\label{proj_G}
\end{equation}
In practice, the frequency-independent coefficients, $a_{k}$,
are obtained through a \GL\ quadrature
(see~\S\ref{sec:LegendreFunctions}).
Equation~\eqref{def_I} then simply becomes
\begin{equation}
M (\omega) = \sum_{k = 0}^{\Ku - 1} a_{k} \, D_{k} (\omega) ,
\label{rewrite_I}
\end{equation}
with
\begin{equation}
D_{k} (\omega) = \!\! \intLb \!\! \rd u \, \frac{P_{k} (u)}{u - \omega} .
\label{def_D}
\end{equation}
Given that the integrand from Eq.~\eqref{def_D}
is analytic,
it is straightforward to apply Landau's prescription
from Eq.~\eqref{Landau_Prescription}
to evaluate the ${ D_{k} (\omega) }$
in the whole complex frequency plane.
We refer to~\S\ref{sec:LegendreFunctions}
for the details of their efficient evaluations.
We also test this implementation in~\S\ref{sec:Plasma}
to recover the damped modes of homogeneous stellar systems.
All in all, problem (B) has been solved.

Equation~\eqref{rewrite_I}
is a great simplification of the difficulty
of the computation of a cluster's response matrix
for ${ \ImPart[\omega] \!<\! 0 }$.
Indeed, the dependence of ${ M(\omega) }$,
w.r.t.\ the cluster's properties
is fully encompassed by the coefficients ${ a_{k} }$
(that must be computed only once),
while its dependence
w.r.t.\ the considered complex frequency,
${ \omega }$, only appears in the analytic functions ${ D_{k} (\omega) }$.

To conclude this section, 
a word is in order to briefly recall
the alternative approach put forward in the seminal work from~\cite{Weinberg1994}.
One can note that
the evaluation of ${ M_{pq}^{\ell} (\omega) }$
from Eq.~\eqref{shape_bM}
does not require any subtle prescription
for ${ \ImPart[\omega] \!>\! 0 }$.
As such, one can evaluate ${ M_{pq}^{\ell} (\omega_{i}) }$
for some given complex frequencies ${ \{ \omega_{i} \}_{0 \leq i \leq 2 d} }$
with ${ \ImPart[\omega_{i}] \!>\! 0 }$.
Noting that Eq.~\eqref{shape_bM}
involves a resonant denominator of the form,
${ 1/(\bn \!\cdot\! \bO (\bJ) \!-\! \omega) }$,
it is then natural to approximate ${ M_{pq}^{\ell} (\omega) }$
with a rational function of the form
\begin{equation}
M_{pq}^{\ell} (\omega) \simeq \frac{R_{pq}^{\ell} (\omega)}{Q_{pq}^{\ell} (\omega)} ,
\label{approx_rat}
\end{equation}
where ${ R_{pq}^{\ell} (\omega) }$ and ${ Q_{pq}^{\ell} (\omega) }$
are polynomials of degree ${ \leq d }$,
inferred from the gridded evaluations.
Because it is analytic,
the rational function from Eq.~\eqref{approx_rat}
can immediately be evaluated in the whole complex frequency plane,
in particular for ${ \ImPart[\omega] \!<\! 0 }$,
i.e.\ to search for damped modes.
In addition, the complex poles of this same approximation
are directly given by the roots of the polynomial denominators.
One obvious advantage of the method from~\cite{Weinberg1994}
is that it does not require the series of change of variables
that led us to Eq.~\eqref{expression_M}.
However, from the numerical point of view,
this alternative method does not really converge
as one increases the density of the nodes $\omega_{i}$ used.
This leads to spurious numerical oscillations
(see, e.g.\@, Fig.~{2} in~\cite{Weinberg1994}),
somewhat similar to the one that are also present
here, e.g.\@, in Fig.~\ref{fig:Plasma},
stemming from the Legendre series truncation.

\section{Application}
\label{sec:Application}

We are now set to apply the previous method
to the case of spherical stellar clusters,
in particular to characterise the properties
of their (weakly) damped ${ \ell \!=\! 1 }$ mode~\citep{Weinberg1994}.
Here, we limit ourselves to the particular
case of the isotropic isochrone cluster,
for which the direct availability of explicit angle-action coordinates
simplifies the numerical implementation.
We refer to~\S\ref{sec:IsochronePotential}
for the associated expressions
and the detailed numerical parameters used throughout.

\begin{figure}
\centering
\includegraphics[width=0.45 \textwidth]{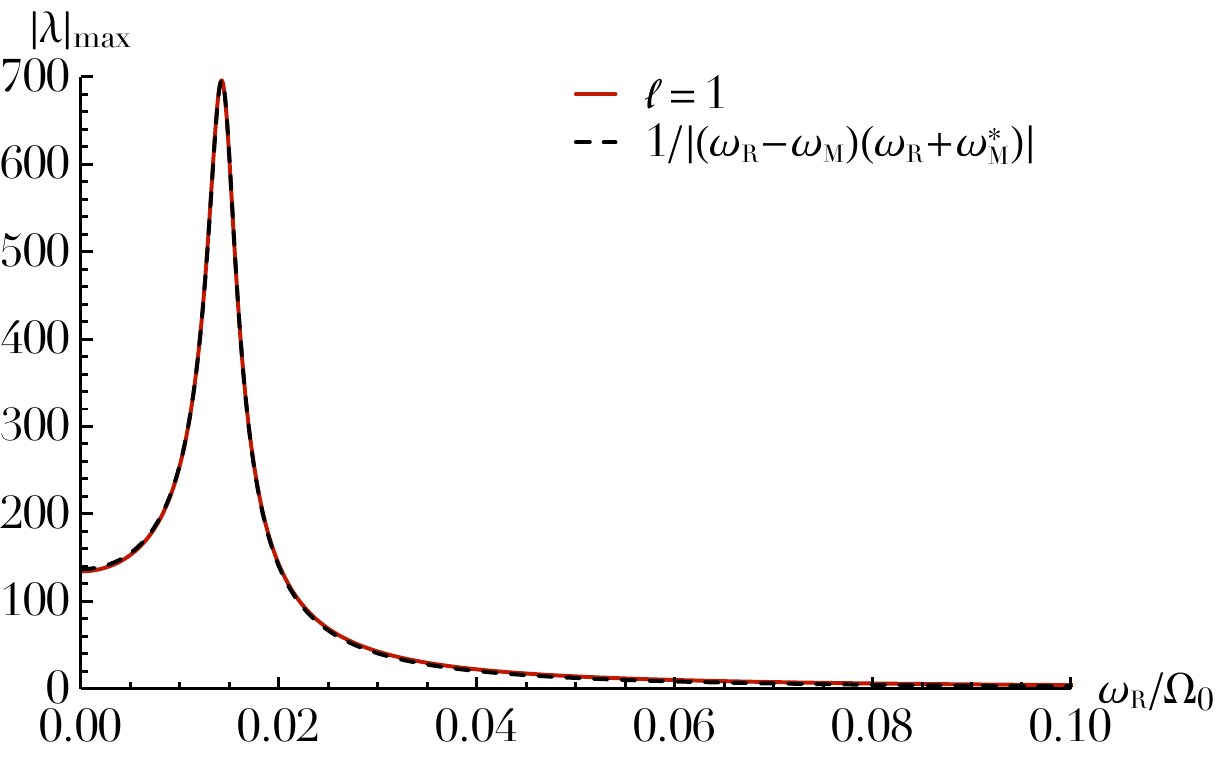}
\caption{Illustration of the maximum eigenvalue norm,
${ |\lambda (\omegaR)|_{\max} }$ of the susceptibility matrix ${ \bN_{\ell} (\omegaR) }$ for ${ \ell \!=\! 1 }$,
as a function of the real frequency ${ \omegaR / \Omega_{0} }$,
with $\Omega_{0}$
the isochrone cluster's frequency scale (see~\S\ref{sec:IsochronePotential}).
One notes a clear amplification for frequencies close
to the real part of the damped mode's complex frequency, ${ \omegaR \!\simeq\! \RePart[\omegaM] }$.
This amplification is well reproduced by a function ${ \!\propto\! 1/|(\omegaR \!-\! \omegaM)(\omegaR \!+\!\omegaM^{*})| }$ that captures the effect
from both damped modes with positive and negative real frequency.
Here, we used the value of $\omegaM$ inferred from Fig.~\ref{fig:Pole},
and only adjusted the function's maximum height.
See \S\ref{sec:IsochronePotential}
for the numerical details.
}
\label{fig:RealLine}
\end{figure}

Before investigating the cluster's ${ \ell \!=\! 1 }$ response matrix
in the lower half of the complex plane,
let us first investigate the cluster's response for purely real frequencies.
This is captured by the susceptibility matrix
\begin{equation}
\bN_{\ell} (\omegaR) = \big[ \bI - \bM_{\ell} (\omegaR) \big]^{-1} ,
\label{def_bN}
\end{equation}
with $\bI$ the identity matrix,
and $\omegaR$ a real frequency.
In Fig.~\ref{fig:RealLine},
we represent the eigenvalue of ${ \bN_{\ell} (\omegaR) }$
that has the largest norm, which we denote
with ${ |\lambda (\omegaR)|_{\max} }$.
In that figure, we clearly observe a narrow amplification in frequency.
This is the direct imprint along the real frequency line
of the cluster's nearby ${ \ell \!=\! 1 }$ damped mode.
Following Eq.~{(139)} of~\cite{Nelson+1999},
it is natural to approximate this narrow amplification with
\begin{equation}
|\lambda (\omegaR)|_{\max} \propto \frac{1}{|(\omegaR - \omegaM) (\omegaR + \omegaM^{*})|} ,
\label{approx_lambda}
\end{equation}
where $\omegaM$ is the complex frequency of the damped mode
(with ${ \RePart [\omegaM] \!>\! 0 }$ and ${ \ImPart [\omegaM] \!<\! 0 }$).
In Eq.~\eqref{approx_lambda},
we also accounted for the contribution from the associated counter-rotating
damped mode, ${ - \omegaM^{*} }$, whose existence is guaranteed
by the cluster's spherical invariance.
In Fig.~\ref{fig:RealLine}, we note in particular
the good agreement
between the approximation from Eq.~\eqref{approx_lambda}
and the numerically measured maximum amplification eigenvalue.

Owing to our explicit analytic continuation of the response matrix
in Eq.~\eqref{rewrite_I},
we may push further this analysis by explicitly evaluating
the cluster's response matrix in the lower half of the complex plane,
i.e.\ for ${ \ImPart[\omega] \!<\! 0 }$.
As such, we define the (complex) dispersion function
\begin{equation}
\veps_{\ell} (\omega) = \det \big[ \bI - \bM_{\ell} (\omega) \big] ,
\label{def_veps}
\end{equation}
so that damped mode corresponds to solutions
of ${ \veps_{\ell} (\omegaM) \!=\! 0 }$ with ${ \ImPart[\omegaM] \!<\! 0 }$.

\begin{figure}
\centering
\includegraphics[width=0.45 \textwidth]{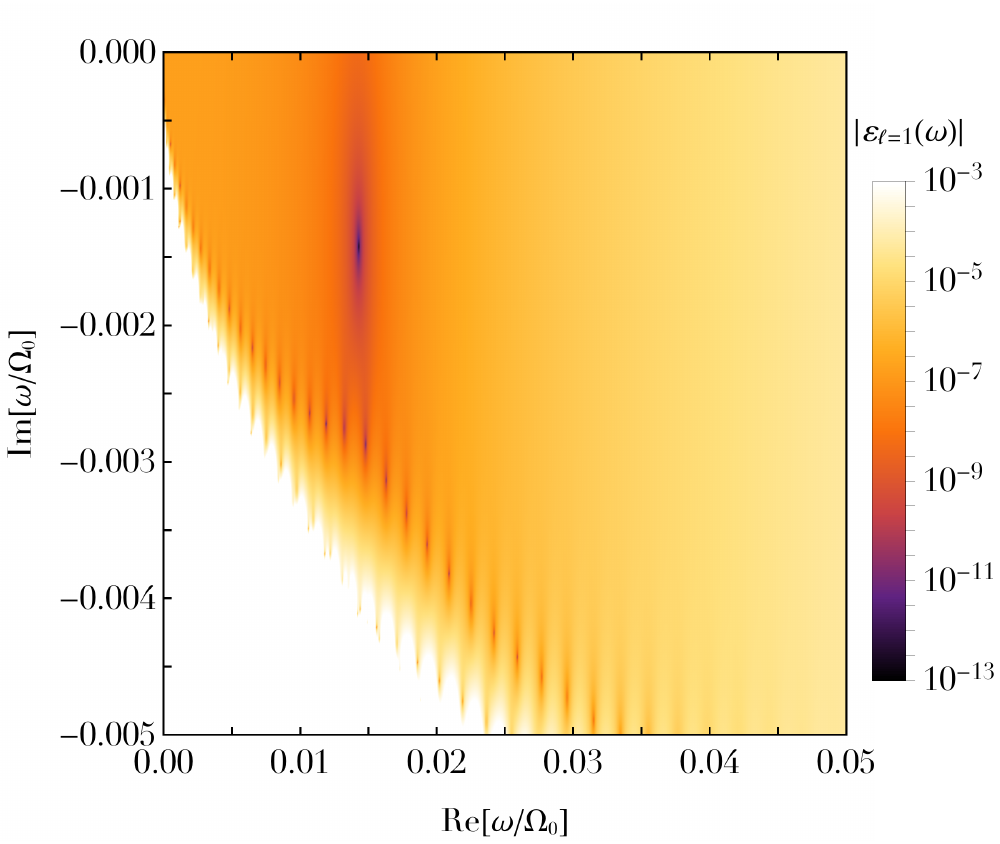}
\caption{Illustration of the ${ \ell \!=\! 1 }$ dispersion function
${ \veps_{\ell} (\omega) }$
in the lower half of the complex plane
for the isotropic isochrone cluster.
One clearly recovers the presence of a (weakly) damped mode
of complex frequency ${ \omegaM/\Omega_{0} \!\simeq\! 0.0143 - 0.00142 \, \ri }$.
Further down in the complex plane,
the present method saturates
and suffers from spurious numerical oscillations.
}
\label{fig:Pole}
\end{figure}

In Fig.~\ref{fig:Pole}, we present this dispersion function
in the lower half of the complex frequency plane.
From that figure, we infer that the isotropic isochrone cluster sustains
a ${ \ell \!=\! 1 }$ weakly damped mode of complex frequency
\begin{equation}
\omegaM / \Omega_{0} \simeq 0.0143 - 0.00142 \, \ri ,
\label{def_oM}
\end{equation}
with ${ \Omega_{0} \!=\! \sqrt{G M / \bISO^{3}} }$
the frequency scale of the isochrone cluster.
We also recall that owing to spherical symmetry,
there exists an associated damped mode
of complex frequency ${ - \omegaM^{*} }$.
We note that the present method
suffers unfortunately from spurious numerical oscillations,
as visible in Fig.~\ref{fig:Pole},
stemming from the Legendre series truncation.

As already pointed out by~\cite{Weinberg1994},
this damped mode, as characterised by Eq.~\eqref{def_oM},
is both
(i) slow since ${ \RePart [\omegaM] \!\ll\! \Omega_{0} }$,
and (ii) weakly damped since
${ |\ImPart [\omegaM] / \RePart [\omegaM]| \!\ll\! 1}$.
These two properties are directly connected since
the fact that ${ \RePart[\omegaM] \!\ll\! \Omega_{0} }$
implies that only a small number of stars
can effectively resonate with the mode
(there are only a few slowly orbiting stars),
which in turn only allows for an inefficient (Landau) damping
of the mode itself.

\begin{figure}
\centering
\includegraphics[width=0.45 \textwidth]{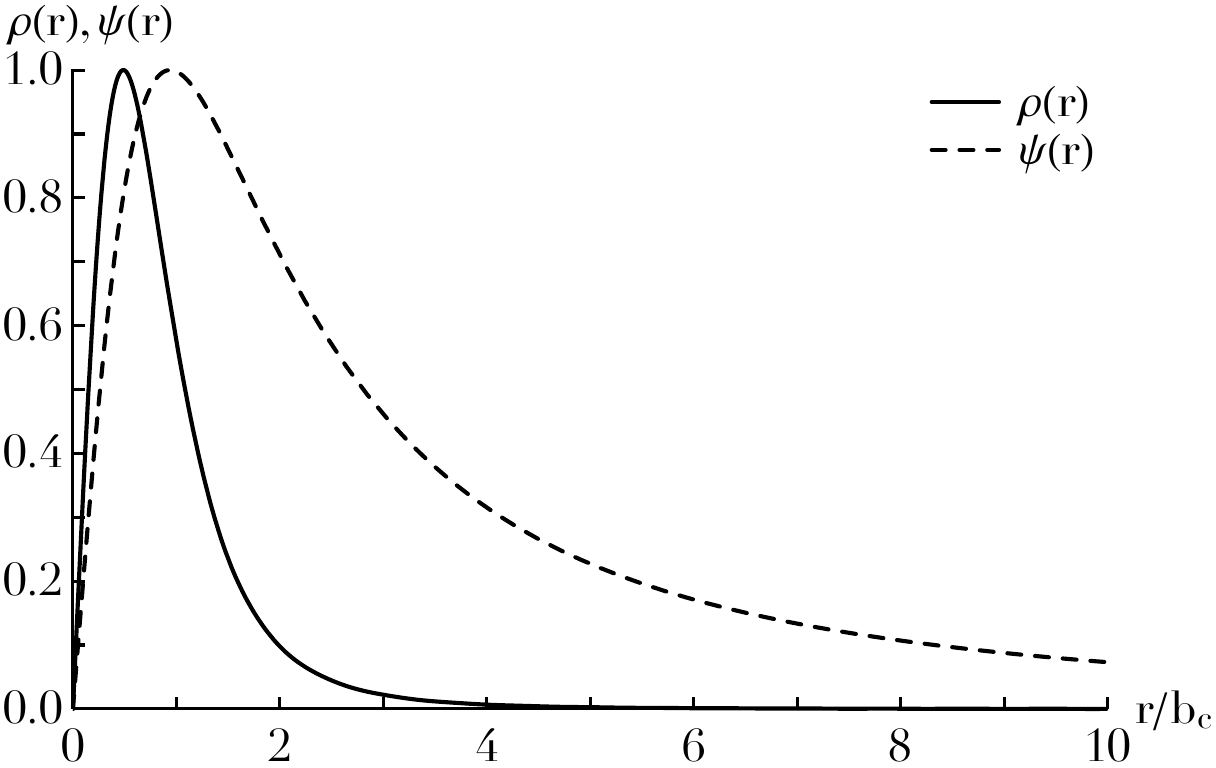}
\includegraphics[width=0.45 \textwidth]{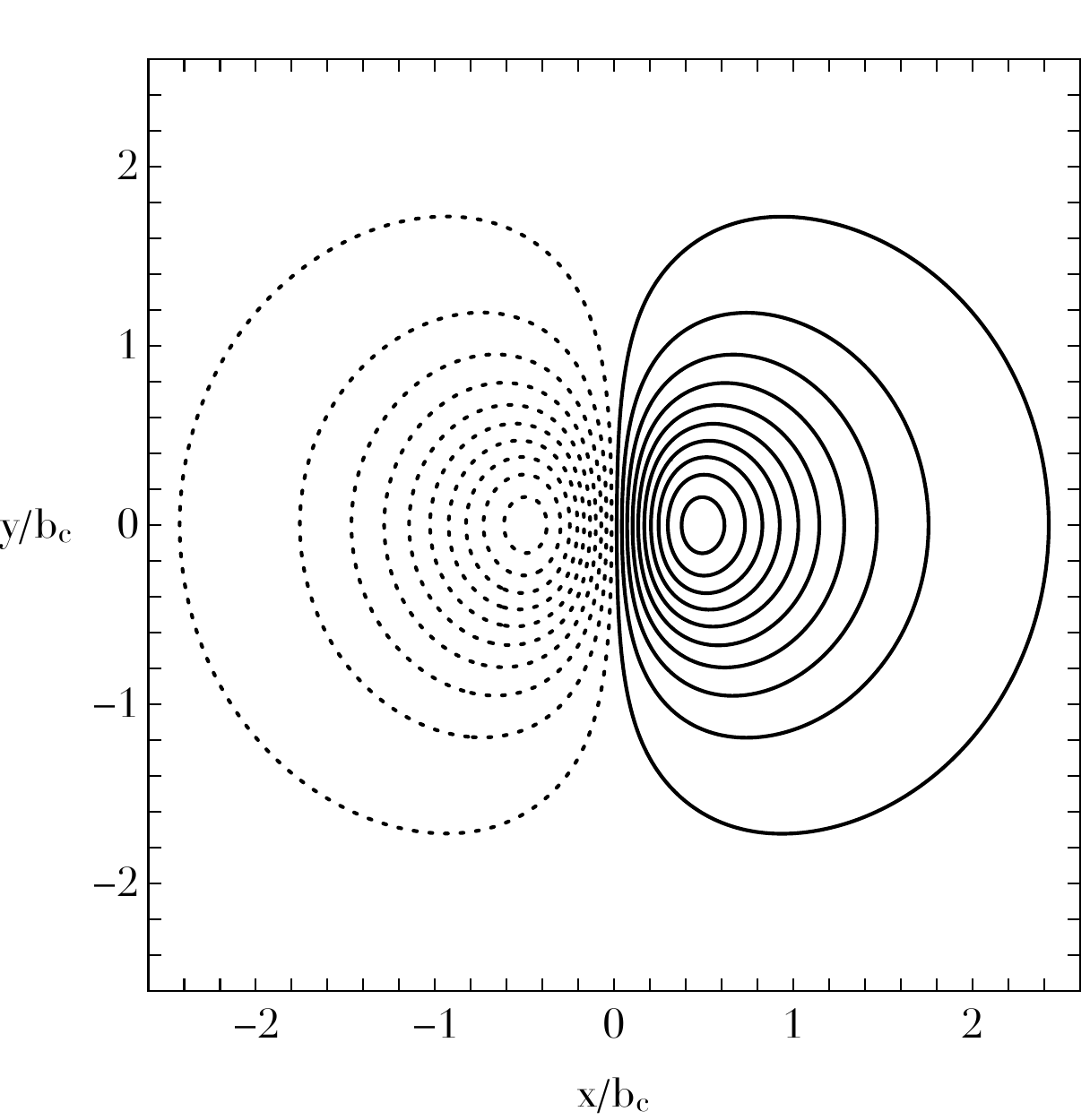}
\caption{Illustration of the density shape
of the ${ \ell \!=\! 1 }$ damped mode
from Fig.~\ref{fig:Pole}.
Top panel: Radial dependence
of mode's density, ${ \rho(r) }$,
and potential, ${ \psi(r) }$.
The normalisation of the vertical axis is arbitrary.
Bottom panel: Illustration of the ${ \ell \!=\! m \!=\! 1 }$
mode's density in the ${(x,y)}$-plane.
Overdensities (resp.\ underdensities) are shown
in solid (reps.\ dashed) lines, spaced linearly
between 95\% and 5\% of the maximum.
}
\label{fig:ShapeMode}
\end{figure}

Having determined the mode's complex frequency,
it is straightforward to obtain the mode's radial dependence.
(see, e.g.\@, Eq.~{(72)} of~\cite{Hamilton+2018}),
as illustrated in Fig.~\ref{fig:ShapeMode}.
In that same figure, we also represent the shape
of the associated ${ \ell \!=\! m \!=\! 1 }$ density perturbation
in the ${(x,y)}$-plane.
We note in particular the striking similarity
with the ${ \ell \!=\! 1 }$ damped mode
already presented in Fig.~{4} of~\cite{Weinberg1994},
in the context of the King sphere
(a cluster with a finite radial extension)
and recovered numerically in Fig.~{6}
of~\cite{Heggie+2020}.

\begin{figure}
\centering
\includegraphics[width=0.45 \textwidth]{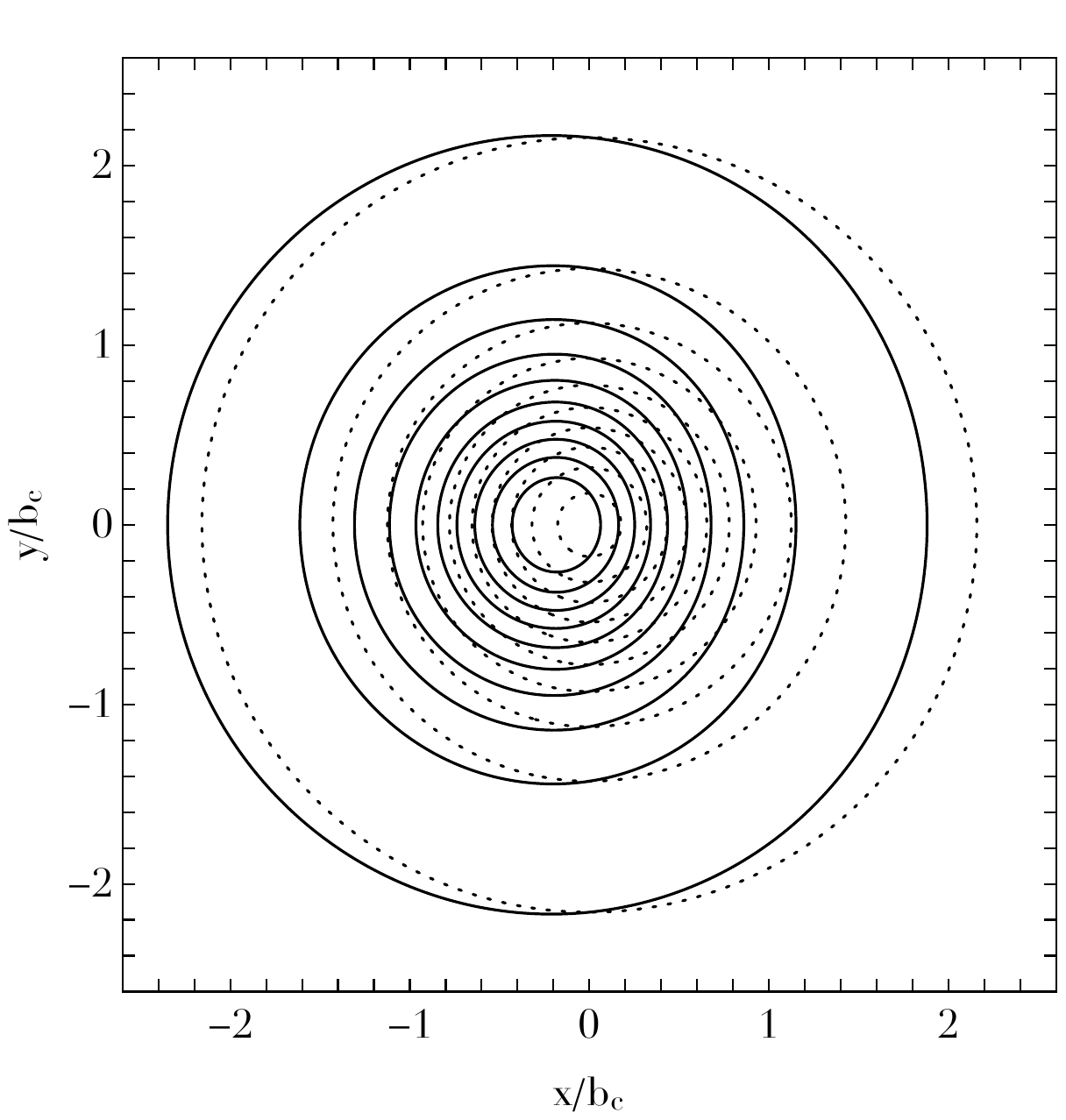}
\caption{Illustration of the overall effect of the density
perturbation from Fig.~\ref{fig:ShapeMode} in solid lines,
with contours spaced linearly between 95\% and 5\%
of the background maximum.
The maximum amplitude of the perturbation is fixed to 20\%
of the background maximum.
The dashed contours correspond
to the unperturbed cluster at the same levels.
As already noted in Fig.~{5} of~\protect\cite{Weinberg1994},
this perturbation leads to a shift of the cluster's centre.
}
\label{fig:ShiftCentre}
\end{figure}

When excited, this ${ \ell \!=\! 1 }$ mode manifests itself
as dipole perturbation.
This leads to a shift
of the cluster's density centre,
as illustrated in Fig.~\ref{fig:ShiftCentre}.
Once again, we emphasise the similarity
with Fig.~{5} of~\cite{Weinberg1994}.
All in all, this drives a sloshing motion of the cluster's centre
of typical frequency ${ \RePart[\omegaM] }$,
which, in the absence of any spontaneous emission,
damps on a timescale set by ${ \ImPart[\omegaM] }$.

While in Fig.~\ref{fig:Pole}
we focused our interest on ${ \ell \!=\! 1 }$ perturbations,
we applied the same method for the other harmonics,
namely ${ \ell \!=\! 0,2,3 }$.
We could not recover any significant damped mode
before falling in the region of spurious numerical modes
(as already visible in Fig.~\ref{fig:Pole}).
Future works should focus on improving
the present scheme's numerical stability
in order to delay the appearance of these artificial modes,
as one explores the lower half of the complex plane
further down.

Having characterised in detail the properties
of the cluster's ${ \ell \!=\! 1 }$ damped mode,
we finally set out to recover its imprints
in direct $N$-body simulations.
We follow an approach similar to the one
presented in~\cite{SpurzemAarseth1996,Heggie+2020},
and we spell out details in~\S\ref{sec:NumericalSimulations}.

\begin{figure}
\centering
\includegraphics[width=0.45 \textwidth]{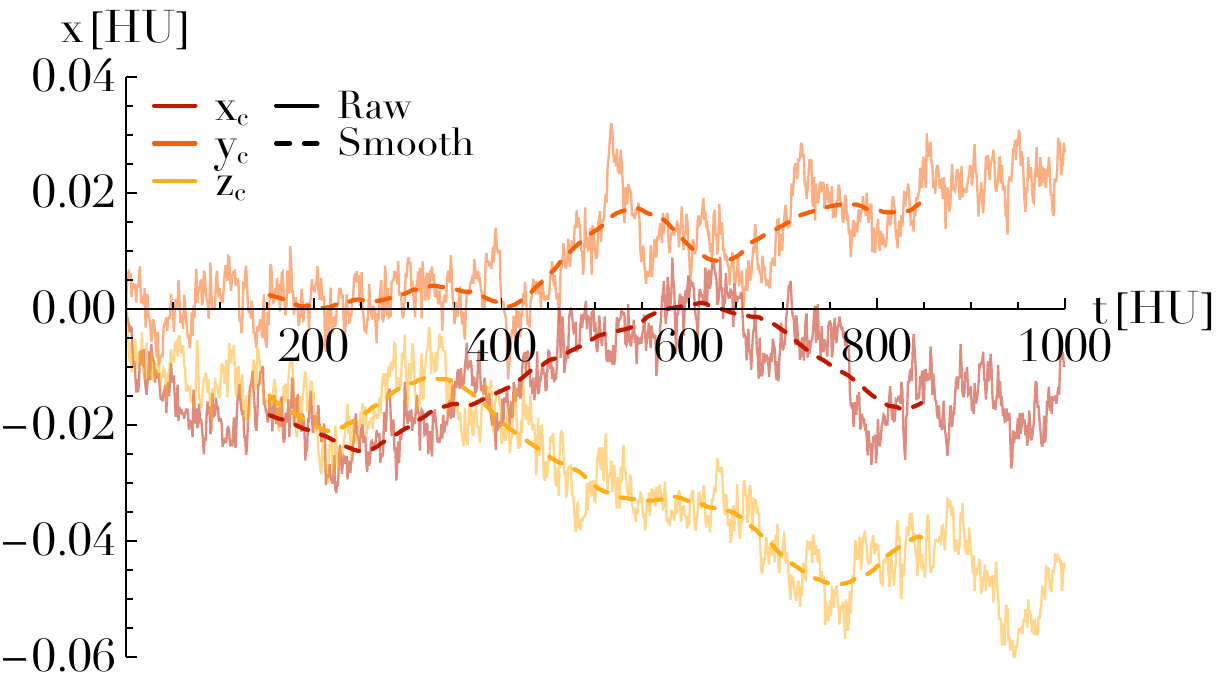}
\caption{Illustration of one realisation of the random walk
of the density centre as measured in one $N$-body realisation,
in H\'enon units ($\HU$; \protect\citealt{Henon1971}).
The dashed lines correspond to the filtered stochastic walks,
see \S\ref{sec:NumericalSimulations}.
The oscillations away from these smooth lines
are notably signatures from the cluster's damped ${ \ell \!=\! 1 }$ mode.
}
\label{fig:NBodyTimeSeries}
\end{figure}

We performed direct $N$-body simulations
of ${ N \!=\! 10^{5} }$ isotropic isochrone clusters
using \texttt{NBODY6++GPU}~\citep{Wang+2015},
keeping track
of the stochastic wandering of the cluster's density
centre~\citep{CasertanoHut1985},
as illustrated in Fig.~\ref{fig:NBodyTimeSeries}
(solid lines).
In order not to be polluted by contributions
from the outermost bound stars (see~\S\ref{sec:NumericalSimulations}),
we filter these time series (dashed lines)
using a \SG\ filter
on a timescale longer than the mode's expected period.
The oscillations of the time series around
their underlying smooth evolutions
are expected to be driven in part by the cluster's damped mode.

To proceed further, similarly to~\cite{Heggie+2020},
we finally estimate the power spectrum, ${ \Pc (\omega) }$,
of the distance of the cluster's density centre
w.r.t.\ its smooth evolution -- see Eq.~\eqref{def_Pc}
for a detailed definition.
Similarly to Eq.~\eqref{approx_lambda},
we expect that for a real frequency $\omegaR$
close to ${ \RePart[\omegaM] }$,
the power spectrum of the density centre
should behave like a Lorentzian of the form
\begin{equation}
\Pc (\omegaR) \propto \frac{1}{|(\omegaR \!-\! \omegaM) (\omegaR \!+\! \omegaM^{*})|^{2}} ,
\label{Lorentzian_Pc}
\end{equation}
where the value of the complex frequency $\omegaM$
was obtained in Eq.~\eqref{def_oM}.
In that expression, similarly to Fig.~\ref{fig:RealLine},
we accounted for the contributions from both damped modes,
i.e.\ $\omegaM$ and ${ - \omegaM^{*} }$,
that only differ in the sign of their real part.

\begin{figure}
\centering
\includegraphics[width=0.45 \textwidth]{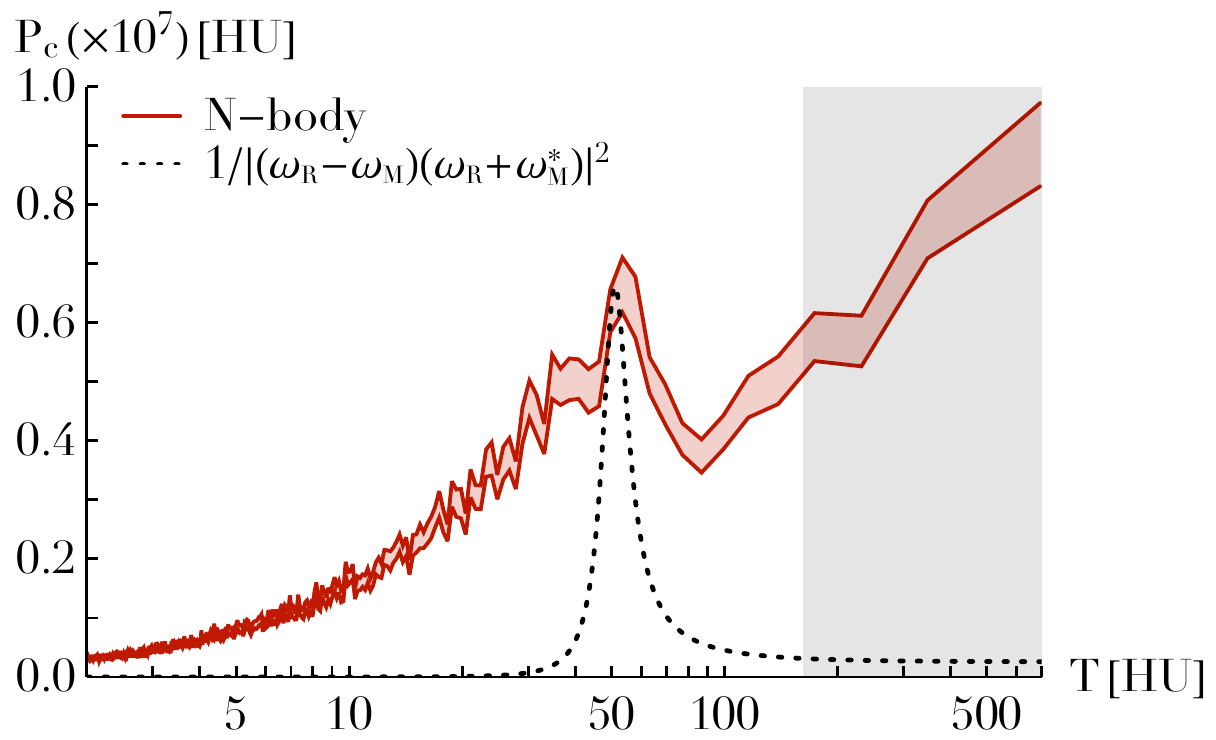}
\caption{Illustration of the power spectrum, ${ \Pc (T) }$,
of the random motion of the cluster's density centre,
as measured in $N$-body simulations (solid lines),
as a function of the period ${ T \!=\! 2 \pi / \omegaR }$
in H\'enon units (${ \HU }$).
Errors are estimated through bootstrap resamplings
over the available realisations,
and the 16\% and 84\% confidence levels are represented.
The dashed line corresponds to the prediction
from Eq.~\eqref{Lorentzian_Pc},
where only the overall amplitude has been adjusted
to the one of the numerically measured peak.
The gray region corresponds to the ${ 3 \, \dB }$ cut region
of the \SG\ filter,
so that it should not be considered.
See \S\ref{sec:NumericalSimulations} for all the numerical details.
}
\label{fig:NBodyCentre}
\end{figure}

In Fig.~\ref{fig:NBodyCentre},
we present the estimation of ${ \Pc (\omega) }$
along with its approximation from Eq.~\eqref{Lorentzian_Pc}.
Reassuringly, we indeed recover the presence
of a narrow amplification peak in the $N$-body simulations
compatible with the one of the ${ \ell \!=\! 1 }$ damped mode
predicted using linear response theory.
However, the measured peak seems
slightly offset and somewhat too wide
compared to the linear theory prediction.
We note that~\cite{Heggie+2020}
(see Fig.~{4} therein),
using a comparable method in King models,
also recovered $N$-body power spectrums
with wider peaks compared
to the linear theory's prediction.
Likely origins for this disagreement include:
(i) an incomplete numerical convergence of the $N$-body simulations;
(ii) non time-stationary signal pollution
associated with the system's initial thermalisation
during which the potential fluctuations get dressed
(see, e.g.\@,~\S{A} in~\cite{Rogister+1968},~\S{F} in~\cite{Fouvry+2018}
and~\cite{Lau+2019}).

\section{Conclusion}
\label{sec:conclusion}

In this work, we have shown how the matrix method
for self-gravitating systems can naturally be tailored
to also compute the (Landau) damped modes of a stellar system.
In order to be able to apply Landau's prescription,
we had to (i) `align' the resonant denominator
with one of the integration coordinate,
and (ii) project the integrand on some explicit basis function
whose analytic continuation is straightforward.
We applied this generic method to the isotropic isochrone cluster
to recover the presence of a low-frequency weakly damped
${ \ell \!=\! 1 }$ mode,
as already demonstrated by~\cite{Weinberg1994}
almost three decades ago.
We also emphasised how the present approach,
because it explicitly separates the orbital dependences
w.r.t.\ the frequency ones (see Eq.~\eqref{rewrite_I})
offers a very efficient numerical scheme.
Finally, following~\cite{Heggie+2020},
we used direct $N$-body simulations to recover
the main properties of this damped mode through the correlated
stochastic motion of the clusters' density centre.
To conclude this work, we now mention
a few possible venues for future works.

For the sake of simplicity, we limited ourselves
to the sole consideration of an isotropic \DF\@, ${ \Ftot \!=\! \Ftot (E) }$,
with an isochrone potential.
Of course, it would be worthwhile to perform
the same investigation for other classes of potentials,
such as King spheres~\citep{Weinberg1994,Heggie+2020}.
We note that applying the current method
to cuspy profiles, e.g.\@, Hernquist,
would require some additional tuning,
since the orbital frequencies, e.g.\@, $\alpha$ as in Eq.~\eqref{def_alpha_beta},
can diverge in the central regions.
Similarly, following, e.g.\@,~\cite{Tremaine2005},
the present method would also require
some further work to be applicable
in dynamically degenerate systems
such as quasi-Keplerian ones.
Finally, lifting the assumption of velocity isotropy,
i.e.\ considering ${ \Ftot \!=\! \Ftot (E , L) }$,
might offer new clues on the numerically observed accelerated relaxation
of tangentially anisotropic clusters~\citep{Breen+2017},
the impact of rotation on a cluster's long-term evolution~\citep{Rozier+2019,Szolgyen+2019,Breen+2021}
or its ${ \ell \!=\! 2 }$ modes.

We also note that our use of the explicitly integrable
isochrone potential eased the numerical tasks
of various steps of the present application:
(i) to perform systematically numerically stable orbit averages
as in Eq.~\eqref{def_W};
(ii) to perform explicitly the change of coordinates
towards the orbital frequencies as in Eq.~\eqref{bM_alpha_beta};
(iii) to determine the range of integrations ${ \vnm (u) }$
and ${ \vnp (u) }$ in Eq.~\eqref{expression_M};
(iv) to easily determine the frequency range probed
by every resonance vector $\bn$,
through ${ \varpi_{\bn} (\omega) }$ in Eq.~\eqref{def_varpi}.
The present method should be extended
to any arbitrary numerically-given radial potential,
along with an appropriately tailored basis expansion~\citep{Weinberg1999}.
Such a generalisation is a necessary first step
to ultimately hope for the explicit time integration
of dressed kinetic equations such as
the Balescu--Lenard equation~\citep[e.g.\@,][]{Fouvry+2021}.

\cite{Lau+2021} recently derived for the first time
the van Kampen modes~\citep{vanKampen1955}
of isotropic stellar clusters, emphasising in particular
how they allow for the detailed characterisation
of the time-stationary thermal fluctuations
present in a stellar cluster.
Following the line of~\cite{Case1959},
it will be undoubtedly be of interest
to fully clarify the connexion between these genuine modes
and the present damped ones.

\cite{Hamilton+2020} recently introduced the quasilinear collision operator
in the context of stellar systems,
to describe the evolution
of the cluster's distribution function as a result of resonant interactions
with Landau damped waves.
Benefiting from the present characterisation
of a cluster's ${ \ell \!=\! 1 }$ damped mode,
one should quantitatively investigate 
the heating signatures arising from such wave-particle interactions.
In particular,
noting that the present ${ \ell \!=\! 1 }$ damped mode
is low-frequency,
so that resonant interactions with the mode only occur
for very small orbital frequencies,
it is expected that this process
will only affect the cluster's outskirts~\citep[see, e.g.\@,][]{Theuns1996},
e.g.\@, through the rate of stellar escapers~\citep{Henon1960}.

Finally, we focused here our interest
on spherical stellar clusters.
Without much work, the present method
could also be applied to razor-thin axisymmetric stellar discs,
since their action space is also ${2D}$.
In particular, it would surely be physically enlightening
to re-interpret the strong swing amplification that inevitably occurs
in sufficiently self-gravitating stellar
discs~\citep[see, e.g.\@,][and references therein]{Binney2020}
as the imprint of (weakly) damped modes.

\subsection*{Data Distribution}
The code and the data underlying this article 
is available through reasonable request to the authors.

\section*{Acknowledgements}

This work is partially supported by grant Segal ANR-19-CE31-0017
of the French Agence Nationale de la Recherche,
and by the Idex Sorbonne Universit\'e.
We thank S.\ Rouberol for the smooth running of the
Horizon Cluster where the simulations were performed,
and G.\ Lavaux for granting access to the GPUs
to run the simulations.
We also warmly thank C.\ Pichon and C.\ Hamilton
for numerous remarks on an earlier version of this work.
JBF is grateful to D.\ Heggie
for his help in designing appropriate $N$-body simulations
and sharing important insights regarding the isochrone cluster.

\appendix

\section{Linear response theory in spherical systems}
\label{sec:LinearTheory3D}

In this Appendix,
we reproduce the key equations
giving the response matrix of spherically symmetric
stellar clusters.
For a spherically symmetric system,
the different spherical harmonics, $\ell$,
decouple from one another.
Following Eq.~{(B9)} of~\cite{Fouvry+2021},
the function ${ G_{pq}^{\ell \bn} (\bJ) }$ introduced in Eq.~\eqref{shape_bM}
reads
\begin{equation}
G_{pq}^{\ell \bn} (\bJ) = - \frac{2 (2 \pi)^{3}}{2 \ell + 1} \big| y_{\ell}^{n_{2}} \big|^{2} \, L \, \bn \!\cdot\! \frac{\p \Ftot}{\p \bJ} \, W_{\ell p}^{\bn} (\bJ) \, W_{\ell q}^{\bn} (\bJ) ,
\label{def_Gpq}
\end{equation}
where we introduced ${ y_{\ell}^{n} \!=\! Y_{\ell}^{n} (\tfrac{\pi}{2} , 0)}$,
with ${ Y_{\ell}^{m} (\vartheta , \phi) }$ the spherical harmonics
normalised so that
${ \!\int\! \rd \vtheta \rd \phi \sin (\vtheta) |Y_{\ell}^{m} (\vtheta , \phi)|^{2} \!=\! 1 }$.
We note that these coefficients impose the constraints
${ |n_{2}| \!\leq\! \ell }$ and ${ (\ell \!-\! n_{2}) }$ even,
and that the vector ${ \bn \!=\! (0,0) }$ does not contribute
to the response matrix.
We also recall that the shape of an orbit is fully characterised
by the action ${ \bJ \!=\! (J_{r} , L) }$,
with $J_{r}$ the radial action, and $L$ the angular momentum,
with the associated orbital frequencies ${ \bO \!=\! (\Omega_{1} , \Omega_{2}) }$.
Equation~\eqref{def_Gpq} also involves the system's
total \DF\@, ${ \Ftot (\bJ) }$,
normalised so that
${ \!\int\! \rd \br \rd \bv \, \Ftot \!=\! M }$
with $M$ the cluster's total mass,
and ${ (\br,\bv) }$ the position and velocity coordinates.

In Eq.~\eqref{def_Gpq},
${ p, q \!\geq\! 1 }$ stand for the radial indices
of the considered biorthogonal basis of potentials and densities.
These are introduced following~\cite{Kalnajs1976},
using the same convention as in Eq.~{(B1)}
of~\cite{Fouvry+2021}.
In practice, owing to spherical symmetry,
it is natural to expand the potential basis elements as
\begin{equation}
\psi^{(\alpha)} (\br) = Y_{\ell}^{m} (\vtheta , \phi) \, U_{n}^{\ell} (r) ,
\label{def_U}
\end{equation}
and similarly for the densities,
with ${ U_{n}^{\ell} (r) }$
some real radial functions
with ${ n \!\geq\! 1 }$.
In practice, we used radial basis elements from~\cite{CluttonBrock1973},
and we refer to~\S{B1} of~\cite{Fouvry+2021} for an explicit expression
of ${ U_{n}^{\ell} (r) }$.
Equation~\eqref{def_Gpq} also involves the coefficients
\begin{equation}
W_{\ell n}^{\bn} (\bJ) = \!\! \int_{0}^{\pi} \!\! \frac{\rd \theta_{1}}{\pi} \, U_{n}^{\ell} (r) \, \cos (n_{1} \theta_{1} + n_{2} (\theta_{2} \!-\! \vphi)) ,
\label{def_W}
\end{equation}
which correspond to the Fourier transform 
of the basis elements w.r.t.\
the canonical angles ${ \bT \!=\! (\theta_{1} , \theta_{2} \!-\! \vphi) }$.
Following~\cite{TremaineWeinberg1984}
(recalled in~\S{A} of~\cite{Fouvry+2021}),
they read
\begin{align}
\theta_{1} = \!\! \int_{\mC} \!\! \rd r \, \frac{\Omega_{1}}{|v_{r}|} ;
\quad
\theta_{2} - \vphi {} & = \!\! \int_{\mC} \!\! \rd r \, \frac{\Omega_{2} - L / r^{2}}{|v_{r}|} ,
\label{AA}
\end{align}
with the radial velocity
${ v_{r}^{2} \!=\! 2 (E \!-\! \psi(r)) \!-\! L^{2}/r^{2} }$
and $\mC$ the contour going from the orbit's pericentre $\rperi$
up to the current position ${ r \!=\! r (\theta_{1}) }$ along the radial oscillation.
Similarly,
${ (\alpha , \beta) }$ from Eq.~\eqref{def_alpha_beta}
are generically given by
\begin{equation}
\frac{1}{\alpha} = \frac{1}{\pi} \!\! \int_{\rperi}^{\rapo} \!\!\!\! \rd r \, \frac{\Omega_{0}}{|v_{r}|} ;
\quad
\beta = \frac{1}{\pi} \!\! \int_{\rperi}^{\rapo} \!\!\!\! \rd r \, \frac{L / r^{2}}{|v_{r}|} ,
\label{alpha_beta_generic}
\end{equation}
with $\rapo$ the orbit's apocentre.
We refer to~\S{B3} of~\cite{Fouvry+2021}
for details regarding the computation of ${ W_{\ell n}^{\bn} (\bJ) }$.
These coefficients are the numerically most demanding quantities.
We finally emphasise that our use of the isochrone potential
allows for straightforward numerically stable angular averages,
as highlighted in Eq.~{(G10)} of~\cite{Fouvry+2021}.

\section{Mapping to orbital frequencies}
\label{sec:OrbitalFrequencies}

In this Appendix, we detail the change of variables
${ (\alpha , \beta) \!\mapsto\! (u,v) }$
used to obtain Eq.~\eqref{expression_M}.

From Eq.~\eqref{def_omegan},
we recall that the resonance frequency,
${ \omega_{\bn} (\alpha , \beta) }$,
is defined as
\begin{equation}
\omega_{\bn} (\alpha , \beta) = n_{1} \, \alpha + n_{2} \, \alpha \, \beta .
\label{def_omegan_repeat}
\end{equation}
For a given $\bn$,
we define the minimum value reached
by ${ \omega_{\bn} }$ as
\begin{equation}
\omega_{\bn}^{\min} \!=\! \Min_{(\alpha,\beta)}[\omega_{\bn} (\alpha , \beta)] ,
\label{def_omegamin}
\end{equation}
and similarly for $\omega_{\bn}^{\max}$,
recalling that ${ (\alpha , \beta) }$ are confined
to the domain from Eq.~\eqref{ranges_alpha_beta}.
We detail in~\S\ref{sec:IsochronePotential}
how $\omega_{\bn}^{\min}$ and $\omega_{\bn}^{\max}$
can easily be determined in the case of the isochrone potential.

We then define the variable $u$ as
\begin{equation}
u  = \frac{2 \, \omega_{\bn} (\alpha , \beta) - \omega_{\bn}^{\max} - \omega_{\bn}^{\min}}{\omega_{\bn}^{\max} - \omega_{\bn}^{\min}} ,
\label{def_u}
\end{equation}
so that ${ -1 \!\leq\! u \!\leq\! 1 }$ by design.
As for the second variable, $v$, we pick
\begin{equation}
v = 
\begin{cases}
\displaystyle \beta & \text{if} \quad n_{2} = 0 ,
\\
\displaystyle \alpha & \text{if} \quad n_{2} \neq 0 .
\end{cases}
\label{def_v}
\end{equation}
With such a choice, the Jacobian of the transformation
${ (\alpha , \beta) \!\mapsto\! (u ,v) }$
simply reads
\begin{equation}
\bigg| \frac{\p (\alpha , \beta)}{\p (u , v)} \bigg| = 
|\omega_{\bn}^{\max} \!- \omega_{\bn}^{\min}| \times 
\begin{cases}
\displaystyle 1 /| 2 n_{1}| & \text{if} \quad n_{2} = 0 ,
\\[2.0ex]
\displaystyle 1/| 2 n_{2} v | & \text{if} \quad n_{2} \neq 0 ,
\end{cases}
\label{Jac_uv}
\end{equation}
where we recall that
${ \bn \!=\! (n_{1} , n_{2}) \!=\! (0,0) }$ does not contribute
to the response matrix.
All in all, this allows us to obtain the expression of ${ G_{\bn} (u,v) }$
from Eq.~\eqref{expression_M} as
\begin{equation}
G_{\bn} (u , v) = \frac{2}{\omega_{\bn}^{\max} \!- \omega_{\bn}^{\min}} \, \bigg| \frac{\p (\alpha , \beta)}{\p (u , v)} \bigg| \, G_{\bn} (\alpha , \beta) ,
\label{exp_G_App}
\end{equation}
with ${ G_{\bn} (\alpha , \beta) }$ introduced in Eq.~\eqref{bM_alpha_beta}.

Let us now detail how the minimum and maximum
frequencies, ${ (\omega_{\bn}^{\min} , \omega_{\bn}^{\max}) }$,
may be determined.
We note that ${ \p \omega_{\bn} / \p \alpha }$
(resp.\ ${ \p \omega_{\bn} / \p \beta }$)
is of constant sign for fixed $\beta$ (resp.\ fixed $\alpha$).
As a consequence,
following Eq.~\eqref{ranges_alpha_beta},
the extrema
of $\omega_{\bn}$ are reached either
in the three edges
${ (0 , \half) }$, ${ (0,1) }$, ${ (1 , \half) }$,
or along the curve ${ (\alpha , \betac (\alpha)) }$
with ${ 0 \!\leq\! \alpha \!\leq\! 1 }$.
This greatly simplifies the search of these extremum values,
as one is left with investigating the behaviour
of the function
\begin{equation}
\omega_{\bn}^{\rc} (\alpha) = n_{1} \, \alpha + n_{2} \, \alpha \, \betac (\alpha) .
\label{def_omegac}
\end{equation}
We detail this characterisation in~\S\ref{sec:IsochronePotential}
for the isochrone potential.

In addition to determining ${ (\omega_{\bn}^{\min} , \omega_{\bn}^{\max}) }$,
Eq.~\eqref{expression_M} also requires
the knowledge of the integration boundaries
${ (\vnm (u) , \vnp (u)) }$.
Following Eq.~\eqref{def_u},
we first introduce the quantity
\begin{equation}
h_{\bn} (u) = \half \big[ \omega_{\bn}^{\max} \!+ \omega_{\bn}^{\min} + u (\omega_{\bn}^{\max} \!- \omega_{\bn}^{\min}) \big] .
\label{def_h}
\end{equation}
For ${ n_{2} \!=\! 0 }$, Eq.~\eqref{def_v}
then gives the simple bounds
\begin{equation}
\vnm (u) = \half ;
\quad
\vnp (u) = \betac (h_{\bn} (u) / n_{1}) .
\label{bounds_v_0}
\end{equation}
For ${ n_{2} \!\neq\! 0 }$, the integration bounds
are more intricate to determine.
Following the mapping from Eq.~\eqref{def_v}
and the allowed domain from Eq.~\eqref{ranges_alpha_beta},
$v$ must satisfy the four constraints
\begin{equation}
0 \!\leq\! v ;
\;\;
v \!\leq\! 1 ;
\;\;
\frac{1}{2} \!\leq\! \frac{h_{\bn} (u)}{n_{2} v} - \frac{n_{1}}{n_{2}} ;
\;\;
\frac{h_{\bn} (u)}{n_{2} v} - \frac{n_{1}}{n_{2}} \!\leq\! \betac (v) .
\label{four_constraints_v}
\end{equation}
The first two constraints are straightforward to account for.
The third one  is also easily incorporated provided one deals
carefully with the respective signs of $n_{2}$
and ${ n_{1} \!+\! \half n_{2} }$.
For ${ n_{2} \!>\! 0 }$,
the fourth and final constraint can be rewritten as
${ h_{\bn} (u) \!\leq\! \omega_{\bn}^{\rc} (v) }$,
with the opposite inequality for ${ n_{2} \!<\! 0 }$.
As a consequence, getting the bounds associated
with this constraint asks for the computation
of the root of the function ${ v \!\mapsto\! h_{\bn} (u) \!-\! \omega_{\bn}^{\rc} (v) }$,
at most twice.
Fortunately, for the isochrone potential (see~\S\ref{sec:IsochronePotential}),
the function ${ \omega_{\bn}^{\rc} (v) }$
is explicitly known, and its extrema as well.
In practice, once appropriate bracketing intervals
are found, we used the bisection method
to find the required roots.
Accounting simultaneously for all these constraints
finally provide us with the values of ${ (\vnm (u) , \vnp (u)) }$
appearing in Eq.~\eqref{expression_M}.
Finally,
for a given value of $u$,
once ${( \vnm (u) , \vnp (u) )}$ determined,
we compute the integral over ${ \rd v }$
in Eq.~\eqref{expression_M}
using a midpoint rule with $\Kv$ nodes.

\section{Isochrone potential}
\label{sec:IsochronePotential}

In this Appendix, we detail
key expressions of the isochrone potential~\citep{Henon1959},
following notations similar to~\S{G} of~\cite{Fouvry+2021}.
We emphasise that the availability of these various analytical
expressions greatly eased the practical implementation
of the method described in the main text.

The isochrone potential is defined as
\begin{equation}
\psi (r) = - \frac{G M}{\bISO + \sqrt{\bISO^{2} + r^{2}}} ,
\label{def_psi_ISO}
\end{equation}
with $\bISO$ the associated lengthscale.
In that case, the frequency scale from Eq.~\eqref{def_Omega0}
takes the simple form
\begin{equation}
\Omega_{0} = \sqrt{\frac{G M}{\bISO^{3}}} .
\label{Omega0_ISO}
\end{equation}
As defined in Eq.~\eqref{def_alpha_beta},
the dimensionless radial frequency, $\alpha$,
and the frequency ratio, $\beta$, take the simple form
\begin{equation}
\alpha = \bigg( \frac{2 E}{E_{0}} \bigg)^{3/2} ;
\quad
\beta = \half \bigg( 1 + \frac{L}{\sqrt{L^{2} + 4 L_{0}^{2}}} \bigg) ,
\label{alpha_beta_ISO}
\end{equation}
with the energy and action scales,
${ E_{0} \!=\! -G M / \bISO }$ and ${ L_{0} \!=\! \sqrt{G M \bISO} }$.
Fortunately, Eq.~\eqref{alpha_beta_ISO} can be readily inverted
to determine the energy and angular momentum of an orbit
with given frequencies.
One simply has
\begin{equation}
E = \half E_{0} \, \alpha^{2/3} ;
\quad
L = L_{0} \, \frac{2 \beta - 1}{\sqrt{\beta (1 - \beta)}} .
\label{inv_alpha_beta_ISO}
\end{equation}
All in all, these mappings allow for the straightforward computation
of the Jacobian ${ |\p \bJ / \p (\alpha , \beta)| }$
appearing in Eq.~\eqref{bM_alpha_beta},
recalling that ${ |\p \bJ / \p (E ,L)| \!=\! 1/\Omega_{1} }$.

As given in Eq.~{(G8)} of~\cite{Fouvry+2021},
along circular orbits, one has
\begin{equation}
\alphac (x) = \bigg( \frac{1}{\sqrt{1 + x^{2}}} \bigg)^{3/2} ;
\quad
\betac (x) = \frac{\sqrt{1 + x^{2}}}{1 + \sqrt{1 + x^{2}}} ,
\label{alpha_beta_circ}
\end{equation}
with ${ x \!=\! r / \bISO }$.
Luckily, these two relations can easily be leveraged
to express the frequency ratio, ${ \betac }$,
along circular orbits, as a function of the associated
(dimensionless) radial frequency.
One gets
\begin{equation}
\betac (\alpha) = \frac{1}{1 + \alpha^{2/3}} .
\label{def_betac}
\end{equation}
It is this function that characterises the bound
of the integration domain in Eq.~\eqref{bM_alpha_beta}.

Finally, for a given resonance $\bn$,
we must determine the minimum and maximum frequencies,
${ (\omega_{\bn}^{\min} , \omega_{\bn}^{\max}) }$
as defined in Eq.~\eqref{def_omegamin}.
This asks for the determination
of the extrema of the function
${ \alpha \!\mapsto\! \omega_{\bn}^{\rc} (\alpha) }$,
as defined in Eq.~\eqref{def_omegac}.
Owing to Eq.~\eqref{def_betac},
for the isochrone potential one has
\begin{equation}
\frac{\p \omega_{\bn}^{\rc}}{\p \alpha} = \frac{P_{\bn} (q)}{3 q^{2}} ,
\label{grad_omegac}
\end{equation}
with ${ q \!=\! 1 \!+\! \alpha^{2/3} }$.
In Eq.~\eqref{grad_omegac}, we also introduced
the polynomial
\begin{equation}
P_{\bn} (q) = 3 n_{1} q^{2} + n_{2} q + 2 n_{2} .
\label{def_Pq}
\end{equation}
Given that ${ P (q) }$ is at most a second-order
polynomial, it is straightforward to determine
the existence/absence of extrema
for the function ${ \alpha \!\mapsto\! \omega_{\bn}^{\rc} (\alpha) }$.
We also note that the same polynomial,
${ P_{\bn} (q) }$, is used to fully characterise
the integration bounds ${ (\vnm (u) , \vnp (u)) }$,
as constrained by Eq.~\eqref{four_constraints_v}.

In order to validate our implementation of the response matrix,
we recovered the radial-orbit instability
in a radially anisotropic isochrone potential,
as investigated in~\cite{Saha1991}.
We refer to \S{G} of~\cite{Fouvry+2021}
for the detailed definition of the considered \DF\@.
For this calculation, we used a total of ${ \nmax \!=\! 100 }$ basis elements
with the scale radius ${ \Rb \!=\! 20 \, \bISO }$
(see \S{B1} in~\cite{Fouvry+2021}),
and the sum over resonance number was truncated at ${ n_{1}^{\max} \!=\! 10 }$.
The orbit-averages in Eq.~\eqref{def_W}
were performed with ${ K \!=\! 200 }$ steps,
while the integrations w.r.t.\ $v$ (see Eq.~\eqref{expression_M})
were performed with ${ \Kv \!=\! 200 }$ steps,
and the \GL\ quadrature used ${ \Ku \!=\! 200 }$ nodes.

\begin{figure}
\centering
\includegraphics[width=0.45 \textwidth]{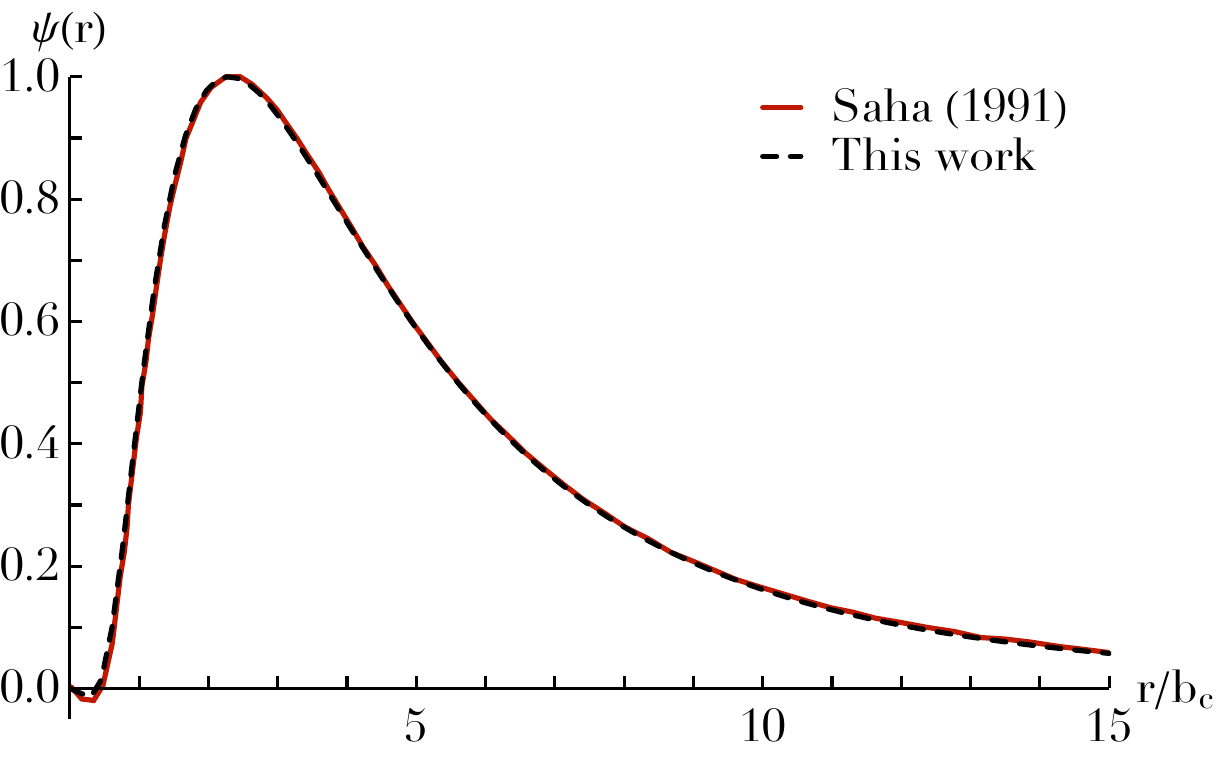}
\caption{Illustration of the radial shape, ${ \psi (r) }$,
as presented in Fig.~{4} of~\protect\cite{Saha1991}
and compared with the present method,
for the ${ \ell \!=\! 2 }$ unstable mode
of a radially anisotropic ${ \Ra \!=\! \bISO }$ isochrone cluster.
The normalisation of the vertical axis is arbitrary.
See the text for the detailed numerical parameters.
}
   \label{fig:Saha}
\end{figure}

In Fig.~\ref{fig:Saha}, we recover that the radially-anisotropic model,
${ \Ra \!=\! \bISO }$ supports an unstable mode with growth rate
${ \eta \!\simeq\! 0.023 \, \Omega_{0} }$ in good agreement with the value
${ 0.024\,\Omega_{0} }$ obtained by~\cite{Saha1991}.
Similarly, the radial shapes of the modes are in good agreement.
This strengthens our confidence in the present method,
at least when searching for instabilities
in the upper half of the complex frequency plane.

\begin{figure}
\centering
\includegraphics[width=0.45 \textwidth]{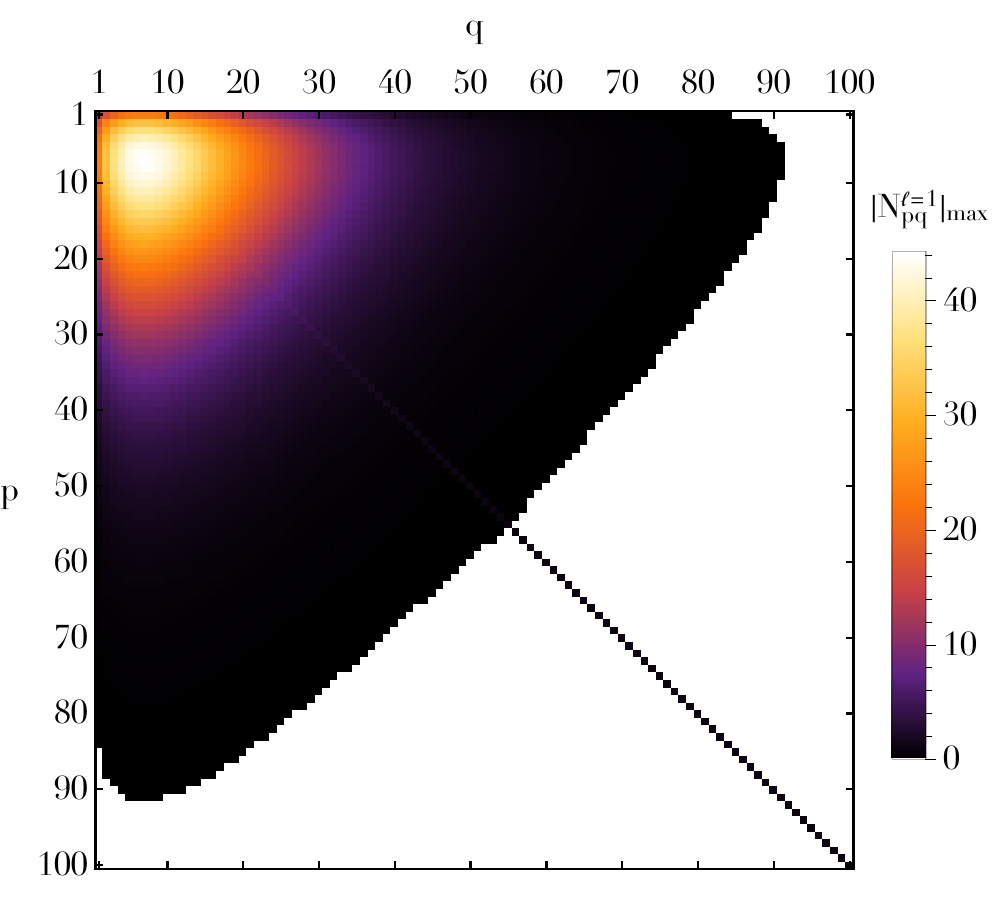}
\caption{Illustration of the maximum norm, ${ |N_{pq}^{\ell}|_{\max} }$,
of the susceptibility matrix ${ \bN_{\ell} (\omegaR) \!=\! [ \bI \!-\! \bM_{\ell} (\omegaR) ]^{-1} }$
for ${ \ell \!=\! 1 }$, as one varies the real frequency $\omegaR$
for fixed ${ (p,q) }$.
Values smaller than ${ 0.05 }$ are replaced with white colors,
highlighting the fact that ${ N_{pq} \!\to\! \delta_{pq} }$
for ${ p , q \!\gg\! 1 }$.
}
\label{fig:NMat}
\end{figure}

In~\S\ref{sec:Application},
when investigating the cluster's ${ \ell \!=\! 1 }$ damped mode,
we used the exact same numerical control parameters
as in Fig.~\ref{fig:Saha}.
In order to further assess the appropriate numerical convergence
of the numerical scheme,
following the same calculation as in Fig.~\ref{fig:RealLine},
we present in Fig.~\ref{fig:NMat}
the maximum norm of the susceptibility matrix
${ | N_{pq}^{\ell} (\omegaR) | }$
as one varies the real frequency $\omegaR$,
for ${ \ell \!=\! 1 }$ and fixed ${ (p,q) }$.
In that figure, we recover in particular
that for ${ p , q \!\gg\! 1 }$,
one has ${ N_{pq} (\omegaR) \!\to\! \delta_{pq} }$,
i.e.\ collective effects can be safely neglected
on small physical scales.

\section{Legendre functions}
\label{sec:LegendreFunctions}

In this Appendix, we detail our use
of the \GL\ quadrature
and our computation of the Legendre functions.

The Legendre polynomials satisfy the normalisation
\begin{equation}
\!\! \int_{-1}^{1} \!\! \rd u \, P_{k} (u) \, P_{\kp} (u) = c_{k} \, \delta_{k \kp} ,
\label{normalisation_Pk}
\end{equation}
with ${ c_{k} \!=\! 2 / (2 k \!+\! 1) }$.
Following Eq.~\eqref{proj_G},
a given coefficient $a_{k}$ is given by
\begin{equation}
a_{k} = \frac{1}{c_{k}} \!\! \int_{-1}^{1} \!\! \rd u \, G (u) \, P_{k} (u) .
\label{calc_ak}
\end{equation}
In practice, these coefficients are directly computed
through a \GL\ quadrature of order $K_{u}$~\citep[see, e.g.\@,][]{Press2007}.
As such, we have at our disposal an explicit set of nodes,
${ \{ u_{i} \}_{1 \leq i \leq K_{u} } }$,
and weights ${ \{ w_{i} \}_{1 \leq i \leq K_{u}} }$.
Then, for any ${ 0 \!\leq\! k \!<\! K_{u} }$,
one approximates the integral from Eq.~\eqref{calc_ak} through
\begin{equation}
a_{k} = \frac{1}{c_{k}} \sum_{i = 1}^{K_{u}} w_{i} \, G (u_{i}) \, P_{k} (u_{i}) ,
\label{calc_ak_explicit}
\end{equation}
noting that the values of ${ \{ P_{k} (u_{i}) \}_{i,k} }$
may be computed once and for all,
independently of the function ${ G (u) }$.

In order to compute the response matrix,
following Eq.~\eqref{def_D},
one has to evaluate the functions
\begin{equation}
D_{k} (\omega) = \!\! \intLb \!\! \rd u \, \frac{P_{k} (u)}{u - \omega} ,
\label{repeat_D}
\end{equation}
The ${ P_{k} (u) }$
are analytic functions that can readily be evaluated
in the whole complex plane.
As such, applying Landau's prescription
from Eq.~\eqref{Landau_Prescription},
we can rewrite Eq.~\eqref{repeat_D} as
\begin{equation}
D_{k} (\omega) =
\begin{cases}
\displaystyle Q_{k} (\omega) & \!\!\! \text{if} \;\; \ImPart[\omega] > 0 ,
\\[2.0ex]
\displaystyle Q_{k} (\omega) \!+\! \ri \pi P_{k} (\omega) \, H(\omega) & \!\!\! \text{if} \;\; \ImPart[\omega] = 0 ,
\\[2.0ex]
\displaystyle Q_{k} (\omega) \!+\! 2 \ri \pi P_{k} (\omega) \, H(\RePart[\omega]) & \!\!\! \text{if} \;\; \ImPart[\omega] < 0 .
\end{cases}
\label{cases_D}
\end{equation}
In that expression, we introduced the Heaviside function
of the ${ [-1,1] }$ interval,
${ H (x) }$ with ${ x \!\in\! \mathbb{R} }$, as
\begin{equation}
H (x) =
\begin{cases}
\displaystyle 0 & \text{if} \quad x < -1 ,
\\[1.0ex]
\displaystyle \half & \text{if} \quad x = -1 ,
\\[1.0ex]
\displaystyle 1 & \text{if} \quad  -1 < x < 1 ,
\\[1.0ex]
\displaystyle \half & \text{if} \quad x = 1 ,
\\[1.0ex]
\displaystyle 0 & \text{if} \quad 1 < x .
\end{cases}
\label{def_H}
\end{equation}
In Eq.~\eqref{cases_D},
we also introduced the function ${ Q_{k} (\omega) }$
defined as\footnote{With the present
convention, for ${ x \!\in\! \mathbb{R} }$
and ${ -1 \!<\! x \!<\! 1 }$,
one has ${ Q_{k} (x) \!=\! - 2 Q_{k}^{\Leg} (x) }$,
with ${ Q_{k}^{\Leg} (x) }$
the usual Legendre function of the second kind.}
\begin{equation}
Q_{k} (\omega) =
\begin{cases}
\displaystyle \!\! \int_{-1}^{1} \!\! \rd u \, \frac{P_{k} (u)}{u - \omega} & \text{if} \quad \ImPart[\omega] > 0 ,
\\
\displaystyle \mP \!\! \int_{-1}^{1} \!\! \rd u \, \frac{P_{k} (u)}{u - \omega} & \text{if} \quad \ImPart[\omega] = 0 ,
\\
\displaystyle \!\! \int_{-1}^{1} \!\! \rd u \, \frac{P_{k} (u)}{u - \omega} & \text{if} \quad \ImPart[\omega] < 0 .
\end{cases}
\label{def_Q}
\end{equation}

The Legendre polynomials, ${ P_{k} (\omega) }$,
generically satisfy Bonnet's recursion formula.
For ${ k \!\geq\! 1 }$, it reads
\begin{equation}
(k + 1) \, P_{k+1} (\omega) = (2 k + 1) \, \omega\,  P_{k} (\omega) - k \, P_{k - 1} (\omega) .
\label{Bonnet}
\end{equation}
Given the definition from Eq.~\eqref{def_Q},
the exact same recurrence relation also applies for ${ Q_{k} (\omega) }$. It now only remains to specify the initial conditions
of these functions.
For the Legendre polynomials, one naturally has
\begin{equation}
P_{0} (\omega) = 1 ;
\quad
P_{1} (\omega) = \omega .
\label{init_P}
\end{equation}
For the function ${ Q_{0} (\omega) }$,
we straightforwardly obtain the expression
\begin{equation}
Q_{0} (\omega) =
\begin{cases}
\displaystyle \ln (1 \!-\! \omega) - \ln (-1 \!-\! \omega) & \text{if} \quad \ImPart[\omega] > 0 ,
\\[2.0ex]
\displaystyle \ln (|1 \!-\! \omega|) - \ln (|-1 \!-\! \omega|) & \text{if} \quad
\ImPart[\omega] = 0 ,
\\[2.0ex]
\displaystyle \ln (1 \!-\! \omega) - \ln (-1 \!-\! \omega) & \text{if} \quad
\ImPart[\omega] < 0 ,
\end{cases}
\label{exp_Q0}
\end{equation}
where the complex logarithm, ${ \ln (\omega) }$,
is defined with its usual branch cut
in ${ \ImPart[\omega] = 0 }$ and ${ \RePart[\omega] < 0 }$.
Finally, noting that
\begin{equation}
\frac{P_{1} (u)}{u - \omega} = \frac{u}{u - \omega} = 1 + \frac{\omega}{u - \omega} = 1 + \omega \, \frac{P_{0} (u)}{u - \omega} ,
\label{pain_de_sucre}
\end{equation}
we can complement Eq.~\eqref{exp_Q0} with the additional relation
\begin{equation}
Q_{1} (\omega) = 2 + \omega \, Q_{0} (\omega) .
\label{exp_Q1}
\end{equation}

\begin{figure}
\centering
\includegraphics[width=0.45 \textwidth]{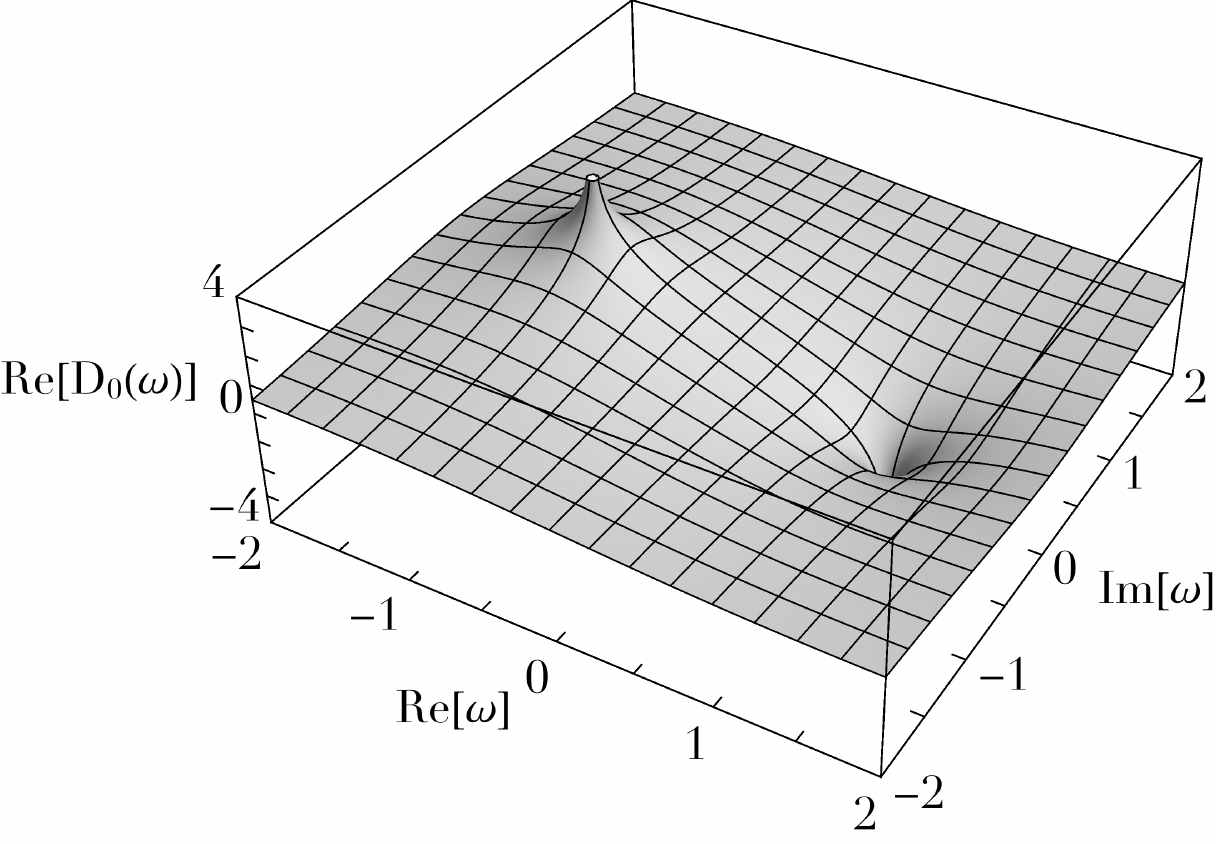}
\includegraphics[width=0.45 \textwidth]{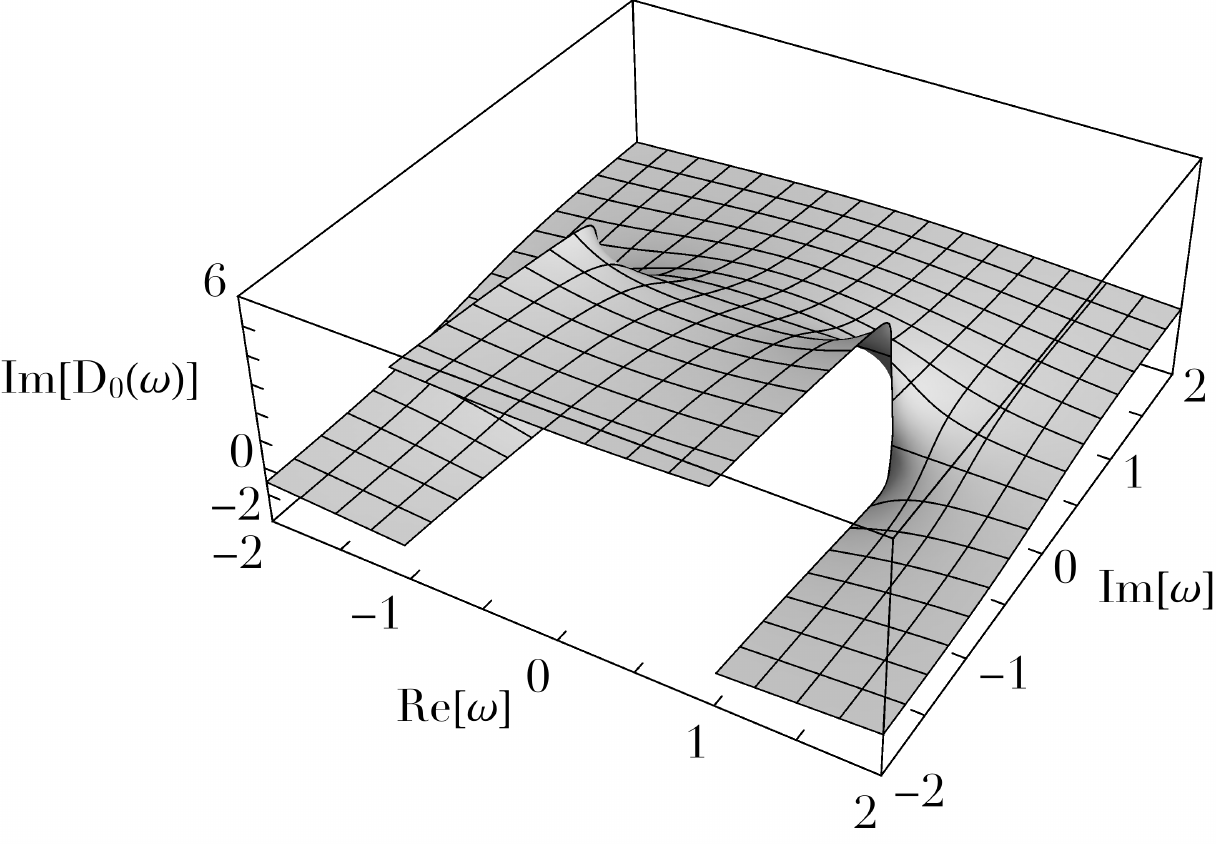}
\caption{Illustration of the complex function ${ D_{0} (\omega) }$
as defined in Eq.~\eqref{repeat_D}.
The top panel corresponds to ${ \RePart[D_{0} (\omega)] }$,
and the bottom one to ${ \ImPart[D_{0} (\omega)] }$.
As expected, this function does not suffer from any
discontinuities in the upper half of the complex plane,
i.e.\ in the region of unstable modes.
}
   \label{fig:Dk}
\end{figure}

In Fig.~\ref{fig:Dk},
we illustrate the behaviour of the function ${ D_{0} (\omega) }$.
We note that this function does not present
any discontinuities in the upper half of the complex plane,
but suffers from two discontinuities
in the lower half, namely:
(i) ${ \RePart[D_{0} (\omega)] }$ diverges in
${ \omega \!=\! \pm 1 }$;
(ii) ${ \ImPart[D_{0} (\omega)] }$ has step discontinuities
along all the lines ${ \ImPart[\omega] \!<\! 0 }$,
in the locations ${ \RePart[\omega] \!=\! \pm 1 }$.
Such discontinuities originate from the fact
that the integral from Eq.~\eqref{repeat_D}
only covers a finite range of frequencies,
i.e.\ ${ -1 \!\leq\! u \!\leq\! 1 }$.

For a given value of $K_{u}$ and a given
complex frequency $\omega$,
a natural way to compute ${ \{ P_{k} (\omega) \}_{0 \leq k < K_{u}} }$
and ${ \{ Q_{k} (\omega) \}_{0 \leq k < K_{u}} }$
is to use Eq.~\eqref{Bonnet}
as a forward recurrence relation.
Namely, one starts from the initial conditions
given by Eqs.~\eqref{init_P},~\eqref{exp_Q0} and~\eqref{exp_Q1},
and, for ${ k \!\geq\! 2 }$, uses the recurrence relation
\begin{equation}
P_{k} (\omega) = \frac{2 k - 1}{k} \, \omega \, P_{k-1} (\omega) - \frac{k - 1}{k} \, P_{k - 2} (\omega) ,
\label{forward_recc_Leg}
\end{equation}
similarly for ${ Q_{k} (\omega) }$.

Yet, for some values of $\omega$ such a recurrence relation
is not numerically stable to compute ${ Q_{k} (\omega) }$.
In that case, we may resort to a backward recurrence.
To do so, we give ourselves a `warm-up' starting point, ${ \Kc \!>\! \Ku }$,
and initialise the recurrence with
\begin{equation}
Q_{\Kc + 2} (\omega) = 0 ;
\quad
Q_{\Kc + 1} (\omega) = 1 .
\label{init_backward}
\end{equation}
Such an initial condition is appropriate
because when the forward recurrence is unstable
it is because one is interested in the decaying mode of recurrence,
which, fortunately, becomes the growing one of the backward recurrence~\citep[see, e.g.\@,][]{Zhang+1996}.
In that case, the recurrence is propagated backwards
using, for ${ k \!\geq\! 0 }$, the relation
\begin{equation}
Q_{k} (\omega) = \frac{2 k + 3}{k + 1} \, \omega \, Q_{k + 1} (\omega) - \frac{k + 2}{k + 1} \, Q_{k+2} (\omega) .
\label{backward_recc_Leg}
\end{equation}
Once ${ Q_{0} (\omega) }$ has been reached,
owing to the linearity of Eq.~\eqref{backward_recc_Leg},
we rescale all the computed values ${ \{ Q_{k} (\omega) \}_{0 \leq k < \Ku} }$
to the correct value of ${ Q_{0} (\omega) }$ given by Eq.~\eqref{exp_Q0}.

For a given value of $\omega$ and $\Ku$,
it only remains to setup a criteria
to specify whether the forward or backward recurrence relation should be used.
In practice, we follow the exact same criteria
as in~\cite{Heiter2010}
(see in particular the function \texttt{qtm1} therein).
The Legendre functions, ${ P_{k} (\omega) }$,
are always computed with the forward recurrence relation
from Eq.~\eqref{forward_recc_Leg}.
For the functions ${ Q_{k} (\omega) }$,
we use the forward recurrence if $\omega$
lies within a given ellipse around the real segment ${ -1 \!\leq\! \RePart[\omega] \!\leq\! 1 }$ and ${ \ImPart[\omega] \!=\! 0 }$.
More precisely, we define
\begin{equation}
b = \Min \bigg[ 1 , \frac{4.5}{(\Ku + 1)^{1.17}} \bigg] ;
\quad
a = \sqrt{1 + b^{2}} .
\label{def_a_b}
\end{equation}
Then, if ever the criterion
\begin{equation}
\bigg( \frac{\RePart[\omega]}{a} \bigg)^{2} + \bigg( \frac{\ImPart[\omega]}{b} \bigg)^{2} \leq 1 
\label{criteria_ellipse}
\end{equation}
is satisfied,
we use the forward recurrence from Eq.~\eqref{forward_recc_Leg}.
When Eq.~\eqref{criteria_ellipse} is not satisfied,
we resort to the backward recurrence from Eq.~\eqref{backward_recc_Leg}.
In that case, the warm-up, $\Kc$ --
see Eq.~\eqref{init_backward}, is determined through
\begin{align}
z {} & = \big| \RePart[\omega] \big| + \ri \, \big| \ImPart[\omega] \big| ;
\quad
c = \sqrt{z^{2} - 1} ;
\nonumber
\\
d {} & = 2 \, \ln (|z + c|) ;
\quad
\Kc = \Ku + \bigg\lceil \frac{\ln(1/\epsilon)}{d} \bigg\rceil ,
\label{calc_Kc}
\end{align}
with ${ \eps \!=\! 10^{-14} }$ a given tolerance target.

\section{Homogeneous stellar systems}
\label{sec:Plasma}

In this Appendix, we apply the method from Eq.~\eqref{rewrite_I}
to an homogeneous stellar
system~\citep[see \S{5.2.4} of][]{BinneyTremaine2008}.
This is useful to test our implementation of the Legendre functions,
as well as the numerical stability of the overall scheme.
As such, we consider the simple case of a $1D$
Maxwellian velocity distribution with
\begin{equation}
M (\omega) = \frac{q}{\sqrt{\pi}} \!\! \intLinf \!\!\!\! \rd u \, \frac{u \, \re^{- u^{2}}}{u - \omega}
\label{def_M_plasma}
\end{equation}
where ${ 0 \!<\! q \!<\! 1 }$
ensures linear stability.
Following Eq.~{(5.64)} of~\cite{BinneyTremaine2008},
one can rewrite Eq.~\eqref{def_M_plasma}
under the simple form
\begin{equation}
M (\omega) = q \big( 1 + \omega \, Z (\omega) \big) ,
\label{rewrite_M_plasma}
\end{equation}
with the usual plasma dispersion function~\citep{Fried+1961}
\begin{equation}
Z (\omega) = \frac{1}{\sqrt{\pi}} \!\! \intLinf \!\!\!\! \rd x \, \frac{\re^{-x^{2}}}{x - \omega} ,
\label{def_Z}
\end{equation}
which can readily be evaluated
in the whole complex plane.

We then compare the analytical expression
from Eq.~\eqref{rewrite_M_plasma} with the result
obtained by applying the method from Eq.~\eqref{rewrite_I}.
To do so, we introduce a truncation velocity, ${ \umax \!>\! 0 }$,
and rewrite Eq.~\eqref{def_M_plasma} as
\begin{equation}
M (\omega) = \!\! \intLb \!\! \rd u \, \frac{G (u)}{u - \varpi} ,
\label{num_M_plasma}
\end{equation}
with
\begin{align}
G (u) = \frac{q \, \umax}{\sqrt{\pi}} \, u \, \re^{- \umax^{2} u^{2}} ;
\quad
\varpi = \frac{\omega}{\umax} .
\label{def_G_plasma}
\end{align}

\begin{figure}
\centering
\includegraphics[width=0.45 \textwidth]{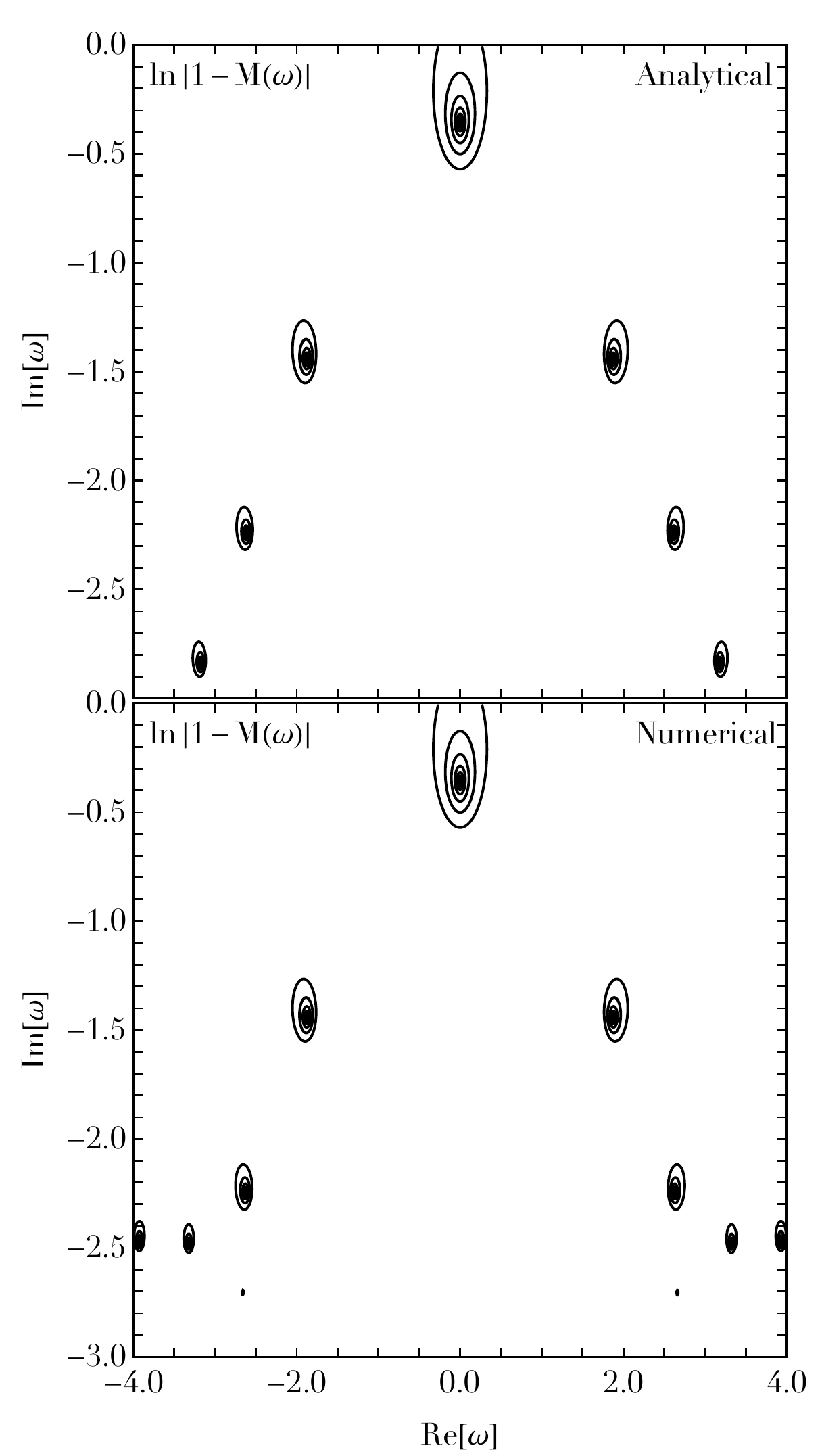}
\caption{Illustration of the contours of the dispersion function,
${ \ln (|\veps (\omega)|) \!=\! \ln (|1 \!-\! M (\omega)|) }$
in the lower half of the complex plane
for the (stable) homogeneous case from Eq.~\eqref{def_M_plasma},
with ${ q \!=\! \half }$.
Around each pole, the outermost contours correspond
to ${ \ln (|\veps|) \!=\! - 0.5 }$ with the subsequent contours spaced every ${ - 0.5 }$.
Top panel: Analytical prediction from Eq.~\eqref{rewrite_M_plasma}
computed with ${ \umax \!=\! + \infty }$.
Bottom panel: Numerical prediction from Eq.~\eqref{num_M_plasma},
truncating the velocity integral at ${ \umax \!=\! 20 }$
and using ${ \Ku \!=\! 200 }$ points to perform
the \GL\ quadrature.
}
   \label{fig:Plasma}
\end{figure}

Given that Eq.~\eqref{def_I} and~\eqref{num_M_plasma}
are of the exact same form,
we may use the same Legendre projection
as in Eq.~\eqref{rewrite_I}.
In Fig.~\ref{fig:Plasma} we illustrate
the contours of the associated dispersion function
in the lower half of the complex frequency space.
We note that both methods are in very good agreement
for the first few damped modes. 
For larger damping rates, the present numerical method
produces spurious zeros,
stemming from the Legendre series truncation --
a problem also encountered in Fig.~{2}
of~\cite{Weinberg1994}.
Similar artificial zeros
are also present in the self-gravitating case
presented here in Fig.~\ref{fig:Pole}.

\section{Numerical simulations}
\label{sec:NumericalSimulations}

In this Appendix, we detail
the numerical simulations used in
Figs.~\ref{fig:NBodyTimeSeries} and~\ref{fig:NBodyCentre}.
The direct $N$-body simulations were performed
with \texttt{NBODY6++GPU}~\citep{Wang+2015}
using a setup identical to the ones presented
in~\cite{Fouvry+2021} (\S{H1} therein).
Each simulation is composed of ${ N \!=\! 10^{5} }$ particles,
integrated for a total duration ${ \tmax \!=\! 10^{3} \, \HU }$,
with an output every ${ \Delta t \!=\! 1 \, \HU }$.
For the isochrone potential,
H\'enon units~\citep{Henon1971}
are such that ${ G \!=\! M \!=\! \Rv }$,
with $M$ the cluster's total mass
and ${ \Rv \!=\! 6 \, \bISO /(3 \pi \!-\! 8) }$ the virial radius.
We performed a total of ${ \Nreal \!=\! 200 }$ different realisations.

For each output, the position of the density centre
was estimated using the algorithm from~\cite{CasertanoHut1985}
with ${ j \!=\! 6 }$ neighbours.
Once the position of the density centre estimated,
we followed the same recentring as in~\cite{Heggie+2020}.
Namely, we place ourselves within the inertial frame
moving along with the system's barycentric uniform motion
and fix the origin of the coordinate's system
so that all the density centres start their evolution from ${ \br \!=\! \bzero }$.
Following these manipulations, each realisation provides us
with three time series, namely ${ \{ \xc (t) , \yc (t) , \zc (t) \} }$.

The density of the isochrone cluster scales likes
${ \rho (r) \!\propto\! 1/r^{4} }$ for ${ r \!\to\! + \infty }$.
As such, it is a rather `puffy' cluster,
i.e.\ one with a significant population
of very loosely bound stars.
More precisely, following Eq.~{(2.51)}
of~\cite{BinneyTremaine2008},
the mass outside radius $r$ in the isochrone model
is of order ${ 2 \bISO M / r }$.
As a consequence, the outermost bound star
is at a radius of order ${ 2 N \bISO }$.
While this outermost star orbits the system,
it will drive significant excursions of the cluster's density centre
of typical lengths ${ 2 \, \bISO \!\simeq\! 0.5 \, \HU }$,
likely visible in Fig.~\ref{fig:NBodyTimeSeries}
(D.\ Heggie, private communication).
We also point out that the timescales associated with 
these oscillations is long,
and cannot be directly resolved within the timespan of our simulations.
In order not to be polluted by these contributions,
one must therefore filter our time series
to better single out the effect of
the cluster's damped mode on the correlated motion
of the cluster's density centre.

In that view, we followed an approach similar
to~\cite{SpurzemAarseth1996}.
For a given realisation, each of the time series,
${ \{ \xc (t) , \yc (t) , \zc (t) \} }$ is filtered
using a \SG\ filter~\citep[see, e.g.\@,][]{Schafer2011}.
Such a filter is characterised by a (half-)window size $\Mf$
and an order $\Nf$.
In practice, in order not to affect the frequency associated
with the damped mode, we used ${ \Mf \!=\! 153 \!\simeq\! 3 \, \TM / \Delta t }$,
with the mode's period ${ \TM \!=\! 2 \pi / \RePart[\omegaM] \!\simeq\! 50.8 \, \HU }$,
as given by Eq.~\eqref{def_oM}.
We arbitrarily fixed the order of the filter to ${ \Nf \!=\! 5 }$.
For a given choice of filtering parameters,
we can then estimate the ${ 3 \, \dB }$ cutoff period of the filter, $\Tc$,
through Eq.~{(12)} of~\cite{Schafer2011}.
It reads here
\begin{equation}
\Tc \simeq \frac{2 \, (3.2 \, \Mf - 4.6)}{\Nf + 1} \, \Delta t \simeq 167 \, \HU .
\label{def_Tcut}
\end{equation}
This cutoff period is such that any signal on period faster
than $\Tc$ is essentially left untouched by the filtering,
hence the requirement ${ \Tc \!>\! \TM }$
to characterise the mode's properties.

Owing to this filtering, we construct
the filtered time series
\begin{equation}
\delta \xc (t) = \xc (t) - \oxc (t) ,
\label{filter_x}
\end{equation}
(similarly for $\yc$ and $\zc$)
where ${ \oxc (t) }$ is obtained via the filtering of ${ \xc (t) }$
and is illustrated in Fig.~\ref{fig:NBodyTimeSeries}.
Following~\cite{SpurzemAarseth1996},
for each realisation we finally construct one time series
\begin{equation}
\delta \Rc (t) = \sqrt{\big[ \delta \xc (t) \big]^{2} + \big[ \delta \yc (t) \big]^{2} + \big[ \delta \zc (t) \big]^{2}} .
\label{def_deltaR}
\end{equation}

Recalling that our effective sampling rate is ${ \Delta t \!=\! 1 \, \HU }$,
each time series consist then of an array
${ \{ \delta R_{i} \}_{0 \leq i < n} }$,
with ${ n \!=\! 1001 \!-\! 2 \Mf }$ the total length
of the available signal after the filtering.
For ${ 0 \!\leq\! k \!<\! n }$, we define the discrete Fourier transform
with the convention
\begin{equation}
\hdR_{k} = \frac{1}{n} \sum_{i = 0}^{n - 1} \delta R_{i} \, \re^{- \ri 2 \pi i k / n} ,
\label{def_DFT}
\end{equation}
so that ${ \hdR_{k} }$ is associated with the frequency
${ \omega_{k} \!=\! \frac{2 \pi k}{n} \tfrac{1}{\Delta t} }$.
Finally, we construct the associated power spectrum
\begin{equation}
\Pc (\omega_{k}) = \big| \hdR_{k} \big|^{2} .
\label{def_Pc}
\end{equation}
This is the quantity presented in Fig.~\ref{fig:NBodyCentre}.

To finalise our measurement,
Eq.~\eqref{def_Pc} is ensemble-averaged
over all the $\Nreal$ realisations available.
In order to estimate the associated errors,
we performed ${ \Nboot \!=\! \Nreal }$
bootstrap resamplings over the available realisations.
This is presented in Fig.~\ref{fig:NBodyCentre}
with the associated ${ 16\% }$ and ${ 84\% }$ error levels.

\balance

\end{document}